\documentclass[twocolumn]{aastex701} %linenumbers, trackchanges
\usepackage{amsmath,amssymb}
\usepackage{bm}

\usepackage{graphicx}
\newcommand{\ds}{\displaystyle}
\newcommand{\xb}{\ensuremath{\boldsymbol{x}}}
\newcommand{\yb}{\ensuremath{\boldsymbol{y}}}

\newcommand{\nb}{\ensuremath{\boldsymbol{n}}}

\newcommand{\Fb}{\ensuremath{\boldsymbol{\mathsf{F}}}}
\newcommand{\Gb}{\ensuremath{\boldsymbol{\mathsf{G}}}}

\newcommand{\Zb}{\ensuremath{\boldsymbol{\mathsf{Z}}}}
\newcommand{\Hb}{\ensuremath{\boldsymbol{\mathsf{H}}}}
 \newcommand{\Phib}{\boldsymbol{\mathsf{\Phi}}}
\newcommand{\eC}{\mathbb{C}}
\newcommand{\eN}{\mathbb{N}}
\newcommand{\eR}{\mathbb{R}}
\submitjournal{ApJS}
\begin{document}
\title{A distributed resource-adaptive implementation of the widefield radio-interferometric measurement model for scalable image formation}
\shorttitle{A distributed resource-adaptive RI measurement model}
\shortauthors{Dabbech \& Wiaux}
\correspondingauthor{Arwa Dabbech}
\author[orcid=0000-0002-7903-3619,gname=Arwa, sname=Dabbech]{Arwa Dabbech} 
\affiliation{Heriot-Watt University, Institute of Sensors, Signal, and Systems}
\email[show]{a.dabbech@hw.ac.uk}

\author[orcid=0000-0002-1658-0121,gname=Yves, sname=Wiaux]{Yves Wiaux} 
\affiliation{Heriot-Watt University, Institute of Sensors, Signal, and Systems}
\email{y.wiaux@hw.ac.uk}

%%%%%%%%%%%%%%%%%%%%%%%%%%%%%%%%%%%%%%%%%%%%%%%%%%

%%%%%%%%%%%%%%%%%%%%%%%%%%%%%%%%%%%%%%%%%%%%%%%%%%

\begin{abstract}
Modern image formation algorithms in radio interferometry rely on repeated applications of the operator $\Phib$ modelling the measurement process and its adjoint $\Phib^\dagger$ to enforce consistency with the acquired data, specifically via their composite mapping $\Phib^\dagger\Phib$ encoding the array's point spread function (PSF). The large data volumes produced during wideband observations yield significant computational challenges for image formation. Moreover, for widefield imaging, the baseline components along the line of sight $w$ complicate severely the measurement model beyond the conventional 2-dimensional non-uniform Fourier transform (NUFFT), making the PSF highly position-dependent. We propose a distributed resource-adaptive implementation of the widefield measurement model, enabled by a hybrid $w$-stacking/$w$-projection approach, whereby the number of $w$-bins is set, in a fully automated manner, to minimise the computational cost under the compute system's memory constraints. The resulting measurement model is naturally decomposed and distributed into low-dimensional operators specific to $w$-bins. Residual $w$-offsets are integrated as measurement-specific Fourier kernels augmenting the sparse de-gridding matrix of the basic NUFFT model. An optional data dimensionality reduction is also introduced, jointly encoding the sequential Fourier de-gridding/gridding operations in $\Phib^\dagger\Phib$ into a holographic matrix when required by memory constraints. For further parallelisation, the sparse de-gridding or holographic matrices are decomposed into blocks via memory-controlled Fourier partitioning. This approach has been validated in prior works through real data case studies for both monochromatic and wideband imaging of MeerKAT and ASKAP data. We provide herein a thorough analysis of its computational efficiency using simulated MeerKAT data. A MATLAB implementation is available in \href{https://basp-group.github.io/BASPLib/}{BASPLib}.

\end{abstract}

%%%%%%%%%%%%%%%%%%%%%%%%%%%%%%%%%%%%%%%%%%%%%%%%%%
%%%%%%%%%%%%%%%%% BODY OF PAPER %%%%%%%%%%%%%%%%%%
\section{Introduction}
From low-frequency radio arrays with all-sky coverage dipoles, such as the Murchison Widefield Array \citep[MWA;][]{Tingay2013} and the Low-Frequency Array \citep[LOFAR;][]{van2013lofar}, to mid-frequency radio arrays with antenna dishes capable of mapping a few degrees in a single pointing, such as the Australian Square Kilometre Array Pathfinder \citep[ASKAP;][]{hotan2021}, and MeerKAT \citep{jonas2016}, modern radio telescopes face significant data processing challenges, particularly pressing in preparation for the data deluge expected from the Square Kilometre Array \citep[SKA;][]{Labate2022,Swart22}. This flagship telescope will target the formation of Petabyte-scale images from Exabyte-scale data volumes \citep{Scaife20}. These challenges arise from both the extreme data acquisition regimes and the complexity of the radio-interferometric (RI) measurement equation. Specifically, widefield observation entails a geometric direction-dependent effect (DDE), termed the $w$-effect, breaking down the assumption that RI data are Fourier measurements of a common sky \citep{Perley1999}. Other sources of DDEs include unknown atmospheric perturbations, and instrumental errors \citep{Smirnov2011}. 

RI image formation approaches, from the standard CLEAN algorithm \citep{hogbom74} and its variants \citep{cornwell08multiscale,Bhatnagar2013}, sparsity-promoting algorithms grounded in convex optimisation theory \citep[e.g.][]{li11,Garsden2015,Dabbech2015} including the SARA family \citep{Carrillo2012,onose2017, Repetti2017,Birdi2019,Repetti2020,Thouvenin2023a}, the Bayesian approaches such as RESOLVE and its extensions \citep{Junklewitz14,Arras21,Roth24}, to the more recent deep learning-based methods \citep{Terris22, Aghabiglou2024,terris2025,Aghabiglou25,Mars25, taja25,tang26}, are iterative in nature. Fundamentally, these algorithms rely on the measurement operator $\Phib$, modelling the measurement process, and its adjoint $\Phib^\dagger$, which are central operators not only for mapping RI Fourier measurements to the image domain via $\Phib^\dagger$ yielding the so-called dirty image, but also to ensure fidelity between the reconstructed image and the observed data within iterative algorithms using $\Phib^\dagger \Phib$. Accurate modelling and computationally-efficient implementations of these operators are critical for the formation of high-resolution high-dynamic range radio images, ultimately achieving the scientific objectives of modern instruments. 

Focusing on widefield imaging, the $w$-effect arises for arrays whose baselines are not coplanar for the duration of the observation, i.e.~their component along the line of sight $w$ is varying. This geometric DDE benefits from a well-defined analytical expression, which in theory, entails a three-dimensional (3D) Fourier transform on the celestial sphere \citep{CornwellPerley1992}. Yet, in practice, encoding the $w$-effect is computationally intractable. Several approaches have been devised to incorporate it efficiently into the measurement model. These are essentially at the interface of two techniques: (i) $w$-stacking \citep{humphreys2011analysis}, and (ii) $w$-projection \citep{Cornwell2008}. On the one hand, $w$-stacking operates directly in the image domain, grouping the $w$ components and their associated RI data on numerous $w$-bins, thus calling for as many 2-dimensional (2D) Fourier transforms \citep[e.g.~][]{Offringa2014,Gheller23}. Fine-sampling of the $w$-bins hampers the scalability of the approach. Whereas, a coarse sampling results in approximate correction, undermining precision. On the other hand, $w$-projection operates in the spatial Fourier domain, encoding the $w$-effect via band-limited convolutional {Fourier} kernels \citep[e.g.~][]{Cornwell2008, Wolz2013, Dabbech2017}. The approach requires a single 2D Fourier transform. However, in extreme fields of view (FoV), $w$-projection kernels present large spatial Fourier bandwidths, requiring significant computational and memory capacity. Hybrid approaches have been explored to balance accuracy and computational efficiency. \citet{Pratley2019} combine $w$-stacking/$w$-projection using fast radially symmetric $w$-projection Fourier kernels in conjunction with $w$-stacking using a small number of $w$-bins. \citet{xie22} combine $w$-stacking/$w$-snapshot, the latter consisting in splitting the data into short time intervals (i.e.~snapshots), within which the array is assumed coplanar. Other approaches such as \citet{Ye21} propose a 3D convolutional de-gridding/gridding framework using least-misfit optimal kernels.

In this article, we present a widefield RI measurement model that addresses the $w$-effect through a resource-adaptive $w$-stacking/$w$-projection approach. The proposed model forms a core component of a framework for RI image formation encompassing iterative state-of-the-art imaging algorithms. These include the monochromatic intensity imaging methods uSARA, a sparsity-based method rooted in optimisation theory \citep{Repetti2020,Terris22}, and its plug-and-play (PnP) counterpart AIRI \citep{Terris22}, as well as their respective wideband extensions Hyper-uSARA and HyperAIRI \citep{tang26}. These algorithms share a forward-backward (FB) iterative structure consisting of a two-step image update: a forward step in the negative gradient direction of a data fidelity function $f$, and a backward step involving a denoising operator associated with either a handcrafted or learned regularisation function $r$. This structure yields a modular imaging framework in which data fidelity and image regularisation are decoupled. Owing to the highly iterative nature of these algorithms, the measurement operator and its adjoint required in the forward step are precomputed once and efficiently distributed across multiple compute nodes throughout the imaging process. The computational complexity of the measurement operator is primarily driven by the number of $w$-bins used for $w$-stacking and the bandwidth of the resulting offset Fourier kernels used for $w$-projection. Our implementation optimises the computational cost under the memory constraints of the compute system. It incorporates three key scalability features: (i) multi-level decomposition of the measurement operator into blocks of sparse operators enabled by a memory-controlled Fourier partitioning; (ii) data dimensionality reduction, which embeds the Fourier de-gridding/gridding operations in a holographic matrix when required by memory limitations; and (iii) a planning strategy for a resource-adaptive implementation with automated selection of the approach to compute the measurement operator including whether to enable data dimensionality reduction, and the associated number of $w$-bins.

The framework is prototyped in MATLAB and has been validated in previous works on real data case studies using the SKA precursor instruments MeerKAT and ASKAP. Early implementations focused on monochromatic image formation via uSARA and AIRI \citep{dabbech22,wilber23a,wilber23b}. Both algorithms demonstrated high-precision imaging by achieving improved resolution and dynamic range compared to the multi-scale variant of CLEAN \citep{Cornwell2008}, implemented in the widely used RI software WSClean \citep{Offringa2014,Offringa2017}, while incurring computational costs within one order of magnitude of CLEAN. More recently, the framework has been validated for wideband image formation using Hyper-uSARA and HyperAIRI, corroborating the trends observed in the monochromatic imaging context when revisiting ASKAP observations of ``the dancing ghosts'', known as PKS 2130-538 \citep{tang26}.

The remainder of this article is organised as follows. Section \ref{sec:rimodel} recalls the RI measurement model in the context of widefield imaging, and the RI image formation problem. A brief overview of the imaging algorithms supported by the proposed RI imaging framework is also provided. Section \ref{sec:scalability} revisits the widefield RI measurement operator model via a hybrid $w$-stacking/$w$-projection approach, and describes the proposed distributed resource-adaptive implementation of the measurement operator and its scalability features. Section \ref{sec:matlab} provides an overview of the MATLAB code library implementing the proposed measurement model. Section \ref{sec:sim} presents validation results based on MeerKAT simulations for monochromatic imaging. Section \ref{sec:realdata} summarises real data case studies using the proposed framework, reported in previous works for both monochromatic and wideband imaging. Section \ref{sec:conclusions} presents the discussion and conclusions.

%-------------------------------------------------------------
\section{Widefield RI image formation}\label{sec:rimodel}
Consider the observation wavelength $\lambda$, the baseline components of any antenna pair in the conventional RI referential system are given by $(u,v,w)\in \eR^3$, where $w$ is the component along the line of sight, and $(u,v)$ represents the projection onto the perpendicular plane. 
Let $(\ell,m)\in \Omega$ denote the direction cosines in the plane tangent to the celestial sphere describing the observed FoV, $\Omega$. RI data, termed visibilities, are expressed as a function of the baselines' components. More specifically, the measurement equation relates the visibility function $\ds{V}_\lambda:\eR^3 \mapsto \eC$ to the sky surface brightness ${\ds I_\lambda}:\Omega\mapsto \eR_+$ such that \citep{Thompson2007}
\begin{equation}
 \label{eq:model-cont}
 {\ds V_\lambda}(u,v;w) =\int_{\Omega} {\ds C}(\ell,m; w) {\ds I_\lambda}(\ell,m) e^{-2i\pi (u\ell+vm)} d\ell dm.
\end{equation}
The function ${\ds C}:\Omega \times \eR \mapsto \eC$ stands for the so-called $w$-term, a chirp-like phase modulation, given by 
\begin{equation}
\label{eq:chrip}
 \ds{C}(\ell,m;w) = e^{-2i\pi w (\sqrt{1-\ell^2 -m^2} -1)}/\sqrt{1-\ell^2 -m^2}.
\end{equation}
In the case of a narrow FoV (i.e.~$ \ell^2+m^2\ll 1 $), 
the $w$-term reduces to $\ds{C}(\ell,m;w)= 1$ for any $w$ and $(\ell,m) \in \Omega$, and the acquired visibilities correspond to Fourier samples of a common radio sky. When mapping a large FoV, the $w$-effect becomes non-negligible. Each antenna pair acquires a Fourier sample of its apparent sky--the target radio sky modulated by its associated direction-dependent $w$-term. Furthermore, visibilities often encompass additional DDEs which can be of various origins, such as atmospheric phase delays, pointing errors, polarisation leakage, to name a few \citep{Smirnov2011}. Unlike the $w$-effect, these modulations are unknown, and require calibration to achieve precision imaging.

Without loss of generality, we focus on the discrete monochromatic RI data model at a given wavelength $\lambda$, as the wideband model follows the same formulation. Neglecting atmospheric and instrumental DDEs, the associated RI data vector is denoted by $\yb \in \eC^M $, and can be formulated as 
\begin{equation}
\label{eq:inverse_pb}
 \yb = \Phib \overline{\xb} + \nb,
\end{equation}
where $\overline{\xb} \in \eR_{+}^N $ is the unknown intensity image of interest, and $\nb\in \eC^M $ is a realisation of a random Gaussian noise with a standard deviation $\tau>0$ and mean 0. The measurement operator $\Phib\in \eC^{M\times N}$ encapsulates both the NUFFT % non-uniform fast Fourier transform 
\citep{Fessler2003}, and the $w$-term. Depending on the science target, often an additional data-weighting scheme accounting for the density of the Fourier sampling \citep[e.g.~Briggs weighting;][] {briggs95} is applied to the data and injected in the measurement operator model for improved resolution. 

Despite their sheer volume, RI data provide incomplete Fourier coverage of the target radio image. The resulting image formation problem is therefore an ill-posed inverse problem. Often, it is addressed via iterative algorithms injecting appropriate image priors into the data. The family of algorithms encapsulated in our framework take the FB iterative structure to impose a handcrafted or a learned image regularisation, while enforcing data fidelity through the function 
 $f(\xb; \yb)=\tfrac{1}{2}\|\Phib\xb-\yb\|_2^2$, which reflects the Gaussian nature of the RI noise, where $\|\cdot\|_2$ denotes the $\ell_2$ norm. Its gradient reads $\nabla f(\xb)=\Phib^\dagger\Phib \xb - \Phib^\dagger \yb$, which, when substituted into the FB gradient step combined with the denoising operator $\operatorname{D}_r$ associated with the image regularisation $r$, leads to the iterative update rule 
 \begin{equation}
 \label{eq:fb}
 \forall k \in \eN, \xb^{(k+1)} = \operatorname{D}_r( \xb^{(k)} +\gamma (\xb_{\textrm{dirty}} - \text{Re}\{\Phib^\dagger\Phib \xb^{(k)}\} )),
 \end{equation} 
where $\gamma$ is a step size. {$\xb_{\textrm{dirty}} \in \eR^N$ is the backprojected data, known as the dirty image and given by $\xb_{\textrm{dirty}}=\mathrm{Re}\{\Phib^\dagger \yb\}$. 
It can be seen that at each iteration the backprojected residual data is obtained from the dirty image by subtracting the contribution of the current image using the composite mapping $\Phib^\dagger\Phib\in \eC^{N\times N}$ encoding the highly position-dependent PSF.} Consequently, these algorithms require multiple passes through the data, raising the need for scalable and accurate models of the measurement process.  

%-------------------------------------------------------------------
\section{Scalable measurement operator model}\label{sec:scalability}
Incorporating the $w$-term in the measurement model via its analytical expression \eqref{eq:chrip} is impractical at scale, as it effectively replaces the FFT with a prohibitively expensive direct Fourier transform applied per measurement. We revisit the widefield measurement operator model using a hybrid $w$-stacking/$w$-projection approach, and propose a distributed resource-adaptive implementation. A comprehensive description of its scalability features is provided. These include: (i) memory-controlled Fourier partitioning enabling multi-level decomposition of the resulting measurement operator, (ii) an optional data dimensionality reduction functionality encapsulating the Fourier de-gridding/gridding operations in a holographic matrix, and (iii) a planning strategy for a resource-adaptive implementation. The resulting measurement operator is formulated in a distributed framework. The overall workflow is summarised for completeness.

\subsection{Revisiting the widefield measurement operator model}
\label{sec:phi}
The geometric DDE can be expressed as $\ds{C}(\ell,m; w) = \ds {C}(\ell,m; w^{\text{stack}}) ~ \ds C(\ell,m; w^{\text{proj}})$, for any $(\ell,m) \in \Omega$, such that $w = w^{\text{stack}} + w^{\text{proj}}$. Formally, the \emph{$w$-stacking} term, $\ds C(\ell,m; w^{\text{stack}})$, represents a phase modulation with a large spatial Fourier bandwidth, encoded directly in the image domain at the spatial resolution of the target radio image, and referred to as the $w$-layer. In contrast, the \emph{$w$-projection} term, $\ds C(\ell,m; w^{\text{proj}})$, accounts for the residual $w$-offset, encoded in the spatial Fourier domain as a compact convolutional 2D kernel with a limited spatial Fourier bandwidth, and referred to as the $w$-kernel.
The gist of $w$-stacking/$w$-projection approach is to discretise the image-sized $w$-terms into a small number of $P$ $w$-layers, such that $P \ll M$, and account for the residual $w$-offsets via $w$-kernels in the spatial Fourier domain. This allows for exact modelling of the $w$-effect within the measurement operator model while ensuring computational efficiency.

\begin{figure}
\centering
 \begin{tabular}{ccc}
 \includegraphics[width=0.17\textwidth]{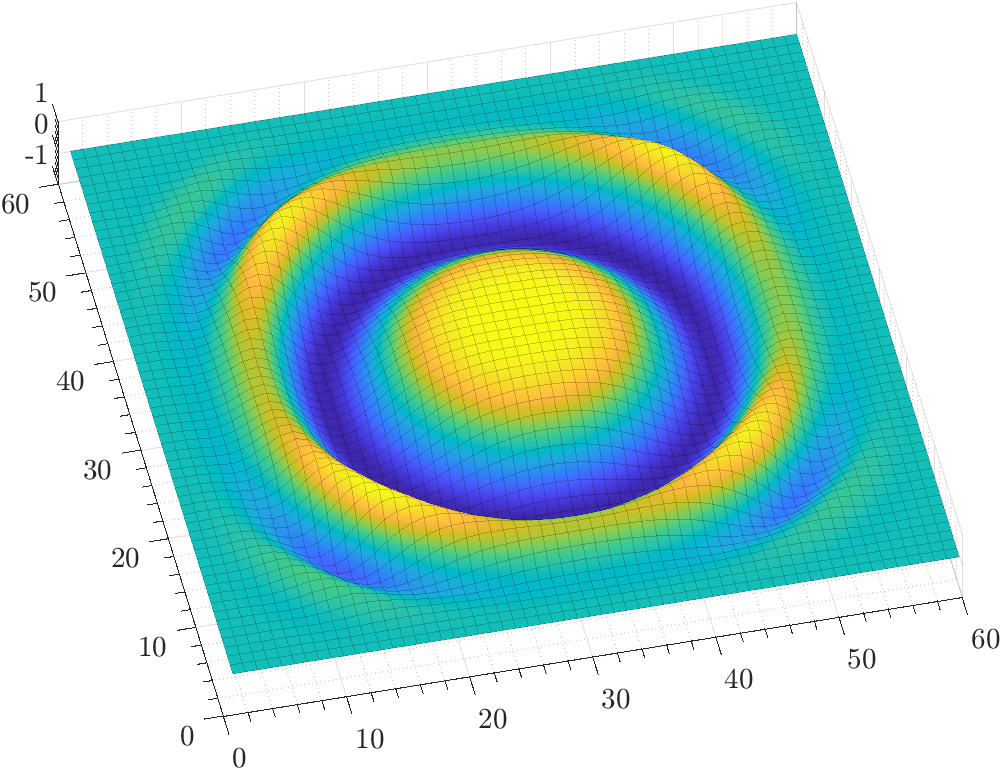}& 
 \includegraphics[width=0.17\textwidth]{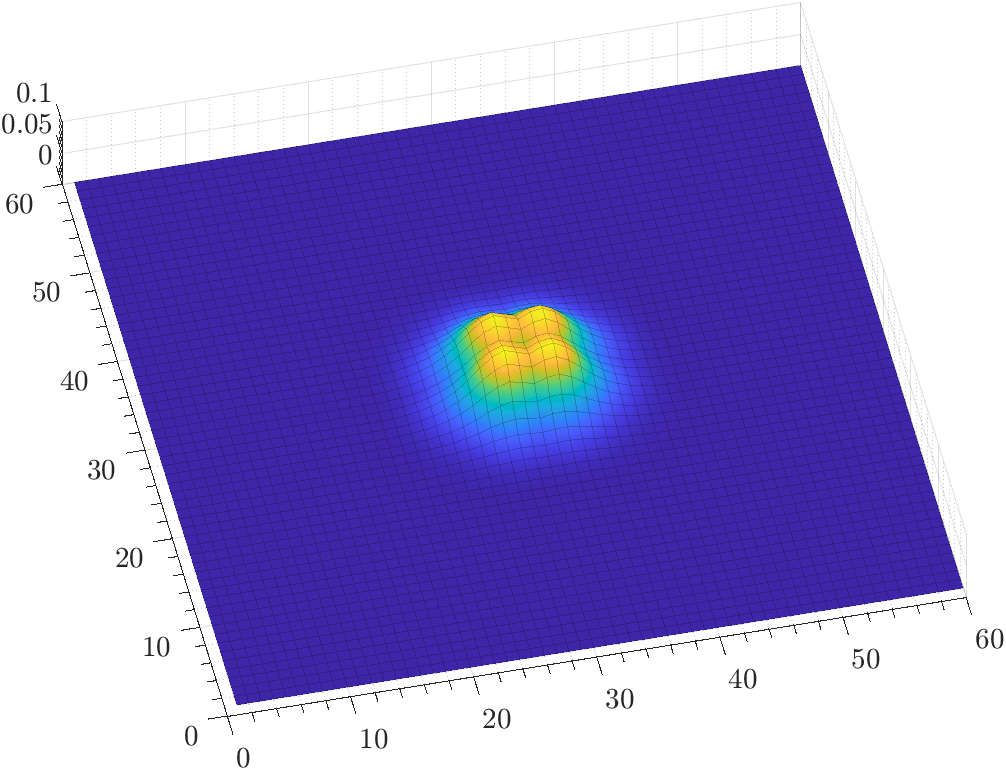} & 
 \includegraphics[width=0.07\textwidth]{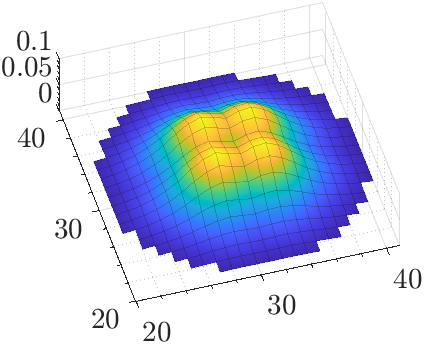} \\
(a) & (b) & (c) 
 \end{tabular}
 \caption{Overview of the approach to obtain the band-limited Fourier $w$-kernels. From left to right: (a) real part of a low-resolution phase modulation, evaluated directly in the image domain over twice the FoV of interest, and tapered via a Hamming window, (b) magnitude of its FFT; (c) magnitude of the $w$-kernel after compression via hard-thresholding.}
 \label{fig:wkernel}
\end{figure}
\paragraph*{$w$-stacking.}~~In principle, the values of the $w$-bins $(w_p^{\text{stack}})_{1 \leq p \leq P} $ should be selected such that the corresponding residual offset components $(w_m^{\text{proj}})_{1 \leq m \leq M}$ yield compact $w$-kernels. Given the non-uniform distribution of the $w$-components, the $k$-means clustering algorithm is employed \citep{Pratley2019}, where identified cluster centroids are adopted as the values of the $w$-bins. To ensure scalability to large data sizes, the clustering is typically performed on a randomly selected subset of the $w$-components. The corresponding $w$-layers, $ \ds C(.~;w^{\text{stack}})$, are then evaluated at the spatial resolution of the target radio image. Of note, in widely-adopted implementations, such as WSClean \citep{Offringa2014}, $w$-bins are sampled uniformly, with the number of bins $P^{\text{WSClean}}$ set to satisfy a phase difference between two subsequent bins below one radian. This condition is ensured via the following lower bound
\begin{equation}
\label{eq:wplanescondt}
P^{\text{WSClean}} \geq 2 \pi \left(w_{\text {max}} -w_{\text{min}}\right) \underset{(\ell,m)\in \Omega}{\mathrm{max}}(1- \sqrt{1-\ell^2-m^2}).
\end{equation}

\paragraph*{$w$-projection.}~~The offsets $(w_m^{\text{proj}})_{1 \leq m \leq M}$ are derived from the corresponding $w$-bins $(w_p^{\text{stack}})_{1 \leq p \leq P} $. To obtain the associated $w$-kernels, the $w$-projection terms $\ds C(.~; w^{\text{proj}})$ are first evaluated at adaptive spatial resolutions, determined from the estimate of their spatial Fourier bandwidth which reads $B(w^{\text{proj}}) = |w^{\text{proj}}|/\Omega$ \citep{Wiaux09}. The $w$-kernels being injected in $\Phib$ directly in the spatial Fourier domain, the sampling of their Fourier grid must be consistent with that of the NUFFT interpolation kernels of $\Phib$ \citep{Dabbech2017}. % (see Section \ref{ssec:dist_op} for details)
The modulations $\ds C(.~; w^{\text{proj}})$ are therefore evaluated over $o=2$ times the size of the FoV of interest. While zero-padding could be applied directly in the image domain, this would be equivalent to the convolution of the Fourier transform of the terms $\ds C(.~; w^{\text{proj}})$ with a sinc function, thereby effectively increasing their spatial Fourier bandwidth. To mitigate this effect, a tapering function, such as a Hamming window, is applied outside the FoV of interest. This ensures that their Fourier transforms remain compact while compensating for the enlarged FoV. The resulting $w$-kernels are then computed via small-sized fast Fourier transforms (FFTs), and further compressed using hard-thresholding based on an $\ell_2$-energy criterion \citep{Wolz2013, Dabbech2017}, preserving a user-defined fraction of the kernel's total Euclidean norm. This approach is illustrated in Figure~\ref{fig:wkernel}. 

Formally, the choice of the number of the $w$-bins $P$ within the hybrid $w$-stacking/$w$-projection approach does not impact the accuracy of the measurement operator. However, it significantly impacts the computational efficiency of its implementation. To address this, we introduce a planning strategy to set the number of $w$-bins $P$ such that the measurement operator fits within the available memory while minimising the computational cost of the resulting measurement operator (see Section~\ref{ssec:planning} for further details).

\subsection{Memory-controlled Fourier partitioning}\label{ssec:dist_op}
The considered hybrid approach to model the $w$-effect enables a natural, first distribution of the widefield measurement operator with respect to the $w$-layers. Considering $P$ $w$-layers, the measurement operator $\Phib$ can be decomposed into $P$ low-dimensional operators $(\Phib_p)$ such that 
% $ where  
\begin{equation}
\label{eq:phi}
\Phib=[\Phib_1,\dots, \Phib_P]^\top, \text{ with } \Phib_p = \Gb_p \Fb \Zb_p~\text{for}~1\leq p \leq P.
\end{equation}
The sparse matrix $\Gb_p \in \eC^{M_p\times N^\prime}$ encodes row-based convolutions between the NUFFT interpolation kernels and the Fourier $w$-kernels. $\Fb \in \eC^{N^\prime \times N}$ denotes the oversampled 2D-Discrete Fourier transform implemented using FFTs, such that $N^\prime = o^2 N$, with $o=2$ denoting the oversampling factor in each spatial dimension. The operator $\Zb_p \in \eC^{N \times N}$ is a diagonal matrix encoding the element-wise product of the $w$-layer ($\ds{C}(.~;w_{p}^{\text{stack}})$), and the so-called grid-correction matrix correcting for the convolution with the NUFFT interpolation kernels. In principle, increasing the number of $w$-layers enables further parallelisation of the measurement operator $\Phib$. However, this requires performing just as many image-sized Fourier transforms, which can limit scalability to large image dimensions. To mitigate this, we introduce a second-level decomposition of the operators ($\Phib_p$), enabled by memory-controlled Fourier partitioning to distribute the computation and application of their underpinning sparse matrices ($\Gb_p)$.

For a given $w$-layer $p$, the associated sparse matrix $\Gb_p$ is split row-wise by partitioning the corresponding $(u,v)$ coordinates via a memory-controlled strategy that adapts to the available compute resources. The memory footprint of each matrix block $\Gb_{p,c}$ is estimated via a lightweight computation of representative rows, modelled as logical kernels which are obtained by row-wise convolution of the corresponding logical $w$-kernel\footnote{In practice, the $w$-components are binned uniformly into a small number of bins to efficiently approximate the representative logical rows.} and a logical NUFFT interpolation kernel. Fourier partitioning is conducted via a radial sweeping of the $(u,v)$-coverage, where a new block indexed by $c+1$ is initiated when the memory limit for the current matrix block $\Gb_{p,c}$ is met. The process is accelerated by splitting the $(u,v)$-coverage into radial chunks for parallel sweeping. The Fourier partitioning also enables the identification of the discrete spatial Fourier coefficients involved in visibility gridding, i.e.~the active columns of $\Gb_{p,c}$. This is critical for allocating sparse matrices with reduced dimension in MATLAB, thereby avoiding the significant memory overhead associated with the default large dimension $N^\prime$. The resulting multi-level decomposition enables full parallelisation of both computation and application of the measurement operator across multiple compute nodes.

\subsection{Holographic matrix for data dimensionality reduction}\label{ssec:datareduct}
{From the iterative rule \eqref{eq:fb} shared by the imaging algorithms supported in our framework, it follows that each FB iteration requires repeated application of the composite mapping $\Phib^\dagger\Phib$, making it the primary operator of interest}, rather than the individual operators $\Phib$ and $\Phib^\dagger$. From the $w$-stacking-driven first-level decomposition, $\Phib^\dagger\Phib$ reads 
{\begin{equation} 
\Phib^\dagger \Phib = \sum^P_{p=1} \Phib_p^\dagger \Phib_p,
\end{equation}
where for each $w$-layer $p \in \{1,\dots,P\}$, the operator
\begin{equation} 
\Phib_p^\dagger \Phib_p = \Zb_p^\dagger \Fb^\dagger \Gb_p^\dagger \Gb_p \Fb \Zb_p, 
\end{equation}
encodes its corresponding position-dependent PSF, which exhibits reduced position dependency compared to that of the full operator.} 
Explicit encoding of the sparse holographic matrices $\Hb_p \equiv \Gb_p^\dagger \Gb_p \in \eC^{N^\prime \times N^\prime}$ would make the operators $(\Phib_p^\dagger \Phib_p)$ blind to the data dimension, and solely dependent on image dimension. Consequently, encoding the Fourier de-gridding/gridding operations through the holographic matrices becomes highly relevant in regimes where the data dimension exceeds the image dimension by several orders of magnitude. Moreover, the holographic matrices are Hermitian by construction. Hence, it is sufficient to store only their lower triangular part, in practice.

A second-level decomposition of the operators $(\Phib_p^\dagger\Phib_p)$ through the distribution of their sparse holographic matrices ($\Hb_p$) into matrix blocks ($\Hb_{p,c}$) can be also enabled following the same approach as for $(\Gb_p)$, detailed in Section~\ref{ssec:dist_op}. However, the memory-controlled Fourier partitioning to distribute $\Hb_p$ can be more computationally demanding. %While the memory footprint of $\Gb_p$ can be easily derived from the estimated spatial bandwidth of its sparse kernels, 
In fact, estimating the memory footprint of the holographic matrix block driven by the number of its non-zero elements, requires the explicit identification of pairs of active discrete Fourier modes in each row of the corresponding de-gridding matrix block. Tracking these pairs during the radial sweep of the $(u,v)$-coverage therefore introduces additional computational overhead. Parallel radial sweeping over chunks of the $(u,v)$-coverage can provide a significant acceleration, although it may lead to an overestimation of memory requirements due to overlap of active Fourier modes at the boundaries of the $(u,v)$-coverage chunks.

\subsection{Compute resource planning}\label{ssec:planning}
To enable a resource-adaptive implementation of the measurement operator, we devise an empirical strategy that uses a small subset of the RI data to determine the computation strategy for $\Phib^\dagger \Phib$, including whether to enable data dimensionality reduction via explicit encoding of the holographic matrix, and the associated number of $w$-layers $P$. The strategy aims to minimise the computational complexity of the operator while ensuring that the cumulative memory footprint of its sparse de-gridding matrices remains within the system's memory budget.

 \begin{figure}
 \centering
 \includegraphics[width=1\linewidth]{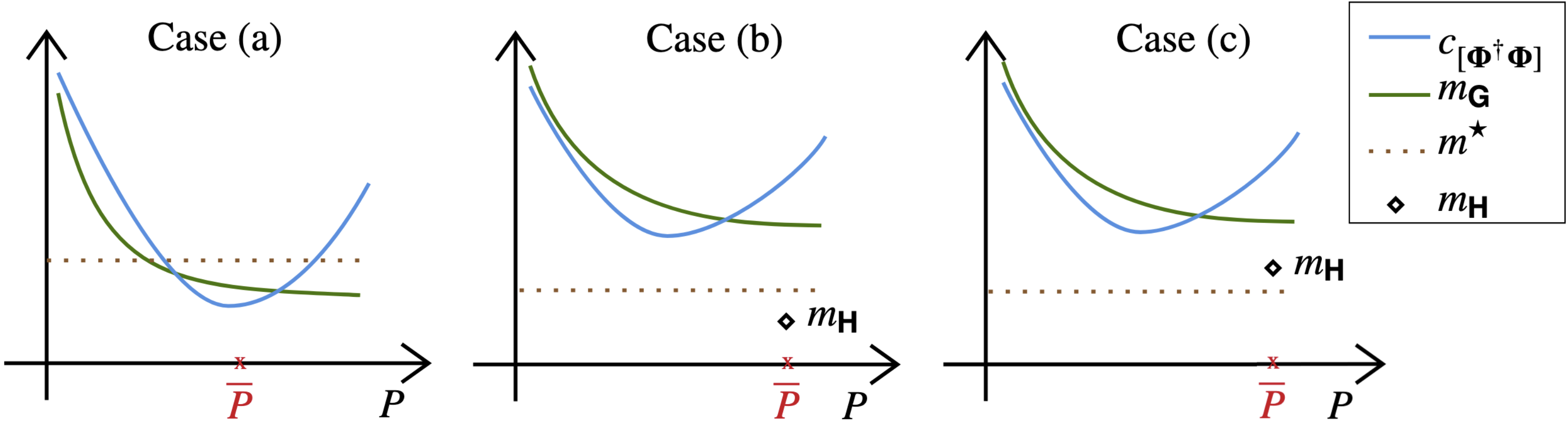}
 \caption{{Illustration of the computational complexity of the operator $\Phib^\dagger \Phib$ (${c}_{[\Phib^\dagger \Phib]}$), and the memory footprint of the underpinning sparse matrix $\Gb$ ($m_{\Gb}$), as a function of the number of $w$-layers ($P$), under the memory budget ($m^\star$). Case (a) corresponds to a regime where $m^\star$ is sufficient, allowing for precomputation of $\Phib^\dagger \Phib$ using $\Gb$. In this case, the selected number of $w$-layers $\overline{P}$ is set to the minimiser of the computational complexity. When $m_{\Gb} > m^\star$ for any value of $P$, the estimation of the memory footprint of the holographic matrix $\Hb$ ($m_{\Hb}$) is triggered at the value $\overline{P}$ at which $m_{\Gb}$ stabilises. Case (b) illustrates a regime where $m_{\Hb}< m^\star$, allowing for precomputation of $\Phib^\dagger \Phib$ using $\Hb$. Case (c) corresponds to a regime in which $m_{\Hb}$ exceeds the available resources, thereby triggering on-the-fly computation and application of ${\Gb}$ when applying $\Phib^\dagger \Phib$ at each iteration of the imaging algorithm. }}
 \label{fig:decisiontree}
\end{figure}

\begin{figure*}
\centering
\includegraphics[width=0.9\linewidth,trim=4.75cm 11.5cm 3.9cm 4.5cm, clip]{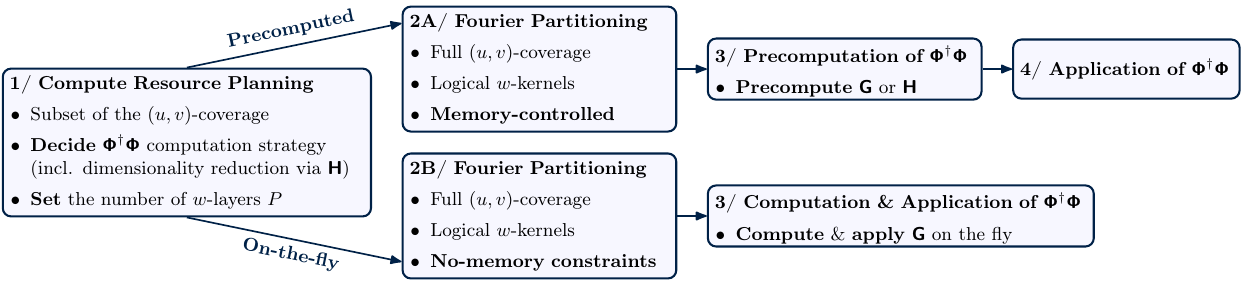}
\caption{{$\Phib^\dagger \Phib$ implementation workflow. 
Resource planning is conducted to determine the operator's computation strategy and associated number of $w$-layers $P$ (stage 1). 
If $\Phib^\dagger \Phib$ is precomputed, memory-controlled Fourier partitioning is performed (stage 2A), followed by precomputation of the underpinning sparse (de-gridding or holographic) matrix blocks (stage 3). The resulting operator is then applied repeatedly within the imaging algorithm (stage 4). If instead $\Phib^\dagger \Phib$ is to be computed and applied on the fly, Fourier partitioning is still conducted to distribute the de-gridding matrix blocks (stage 2B) for parallel on-the-fly computation and application using all available workers within the imaging algorithm (stage 3). 
}}
\label{fig:workflow}
\end{figure*}

Let $m_{\Gb}$ denote the memory footprint of the sparse de-gridding matrix $\Gb=[\Gb_1, \dots,\Gb_P]^\top$, with $K$ the number of its non-zero elements. Increasing the number of $w$-layers $P$ results in smaller $w$-offsets and narrower spatial bandwidths of the corresponding $w$-kernels, thereby reducing $m_{\Gb}$. Conversely, small values of $P$ lead to larger $w$-offsets, which result in $w$-kernels with increased spatial support and a correspondingly denser matrix $\Gb$, thereby increasing $m_{\Gb}$. Expressing $K$ as a function of $P$, the memory required to encode $\Gb$ in MATLAB is given by
\begin{equation}
 m_{\Gb}(P) = 24 K(P) + 4N^\prime. 
\end{equation}
In the asymptotic regime where $P$ is large, the $w$-offsets become negligible, $\Gb$ reduces to the NUFFT interpolation matrix, with ${K} = M Q$, where $Q$ represents the support of the NUFFT interpolation kernel, and $m_{\Gb}$ converges to $m^\star_{\Gb} = 24 MQ$.

The computational complexity of $\Phib^\dagger \Phib$, denoted as ${c}_{[\Phib^\dagger \Phib]}$, is determined by the number of image-sized FFTs and the matrix-vector multiplications underlying the Fourier de-gridding/gridding operations, i.e.~when applying $\Gb^\dagger \Gb$. It can be estimated as 
\begin{equation}
\label{eq:computecomplexity}
 {c}_{[\Phib^\dagger \Phib]}(P) \approx 2(10 N^\prime \log_2 \sqrt{N^\prime} P+ 6 K(P)).
\end{equation}
For large $P$, the complexity is dominated by the increasing number of FFTs to be performed, whereas for small $P$ it is dominated by the sparse matrix-vector multiplications in the Fourier de-gridding/gridding operations owing to the increased density (i.e.~number of non-zero elements) of $\Gb$. 

If there exists a value of $P$ for which the memory footprint of $\Gb$ can be accommodated within the available memory budget $m^\star$, the operator $\Phib^\dagger \Phib$ is precomputed using $\Gb$. Let $\overline{P}$ denote the selected number of $w$-layers, and $P^\star$ denote the minimiser of ${c}_{[\Phib^\dagger \Phib]}$. If $m_{\Gb}(P^\star) < m^\star$, the number of $w$-layers is naturally set to $\overline{P} = P^\star$. Otherwise, $\overline{P}$ is set to the closest value to $P^\star$ such that $m_{\Gb}(\overline{P}) < m^\star$. When $m_{\Gb}$ stabilises above the memory budget $m^\star$, data dimensionality reduction via the holographic matrix is considered.

Estimating $m_{\Hb}$, the memory footprint of the holographic matrix $\Hb = [\Hb_1, \dots, \Hb_P]^\top$, is inherently more challenging than $m_{\Gb}$. Therefore, dimensionality reduction is investigated at a single value $\overline{P}$ that corresponds to the regime in which $m_{\Gb}$ stabilises, ensuring a reduced number of active discrete Fourier modes in the columns of $\Gb$ and, consequently, a lower $m_{\Hb}$. If $m_{\Hb}$ remains within the memory budget, $\Phib^\dagger \Phib$ is precomputed using $\Hb$. Otherwise, its precomputation becomes unfeasible, and its computation and application are performed on the fly at each iteration of the imaging algorithm. In this regime, the cost of $\Phib^\dagger\Phib$ application at each iteration is fully dominated by the computation of the de-gridding matrix blocks, and $\overline{P}$ is therefore set to the value at which $m_{\Gb}$ stabilises. Representative scenarios triggering the three strategies for $\Phib^\dagger \Phib$ computation and the number of $w$-layers selected accordingly is provided in Figure~\ref{fig:decisiontree}.

Finally, since each $w$-layer is assigned its own pool of workers, the number of CPU cores required is lower-bounded by the number of $w$-layers. Satisfying memory constraints alone is insufficient; the allocated resources must provide a sufficient number of CPU cores.

\subsection{Distributed \& parallel application of \texorpdfstring{$\Phib^\dagger \Phib$}{PhidagPhi}}
\begin{figure*}
\includegraphics[width=.96\linewidth,trim=1.cm 5.25cm 2.6cm 1.cm, clip]{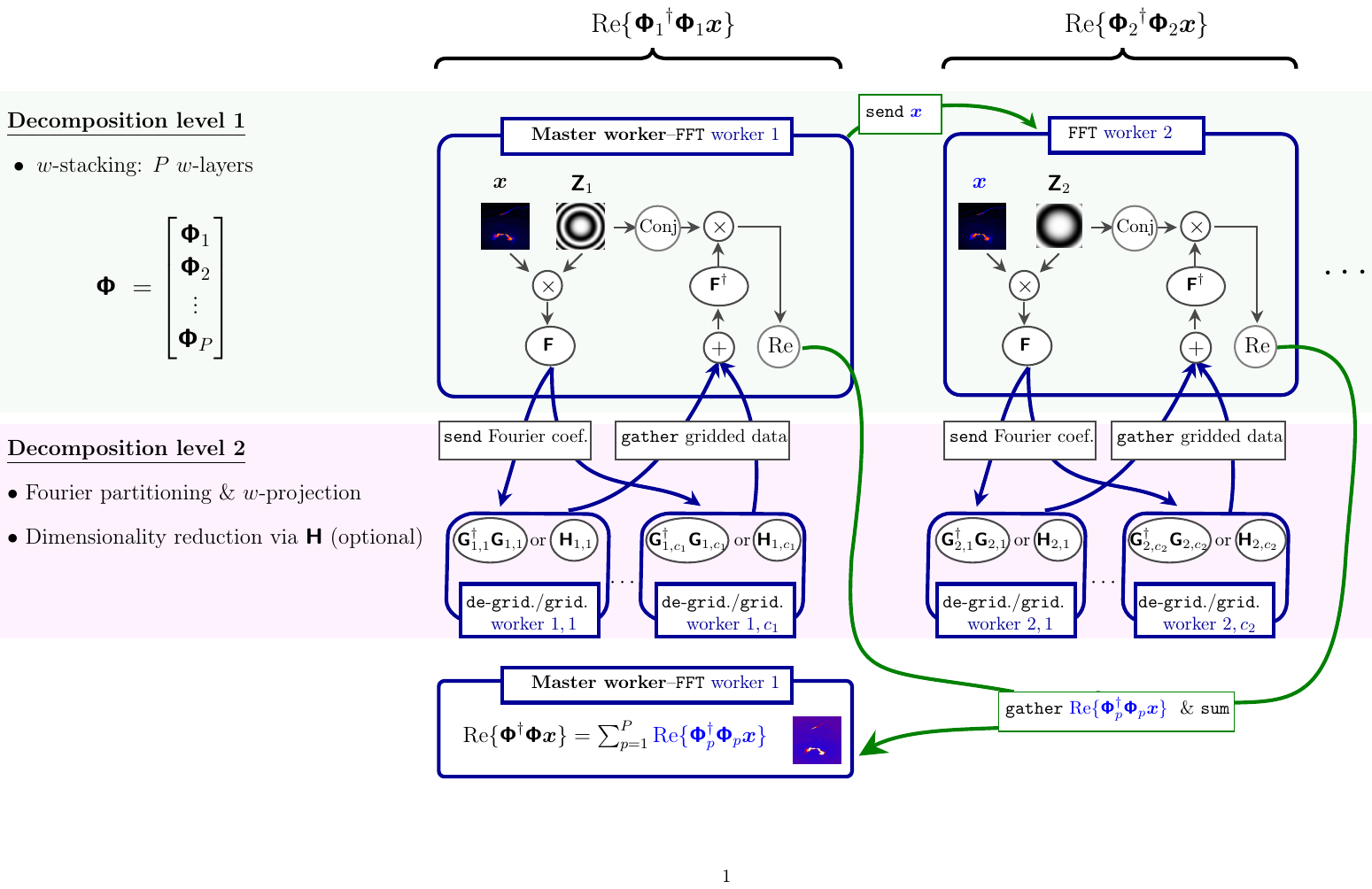}
\caption{{Application of the operator $\Phib^\dagger \Phib$ to an input image ${\xb}$ hosted by the master worker. Two levels of decomposition are considered for full parallelisation. The first level arises from the $w$-stacking, whereby $P$ FFT workers--corresponding to the $w$-layers--perform Fourier transforms (specifically, the operators $\Fb \Zb_p$). The second level, enabled by a memory-controlled Fourier partitioning, distributes de-gridding/gridding operations $\Gb_p^\dagger \Gb_p$ (or $\Hb_p$ when dimensionality reduction is enabled) across the de-gridding/gridding workers. The images $\text{Re}\{\Phib^\dagger_p \Phib_p {\xb}\}$ are gathered from FFT workers in the master worker and are summed to obtain $\text{Re}\{\Phib^\dagger \Phib {\xb}\}$. Cross-worker communication consists of (i) transfer of image-sized variables to and from FFT workers (green arrows), and (ii) tranfer of Fourier coefficients to and from de-gridding/gridding workers (blue arrows). Cross-worker communication follows a two-branch tree structure to reduce overhead and idle time and enable efficient data exchange.}} 
\label{fig:measop}
\end{figure*}

We provide details of $\Phib^\dagger \Phib$ application within an iterative imaging algorithm, as illustrated in Figure~\ref{fig:measop}. 
It follows the two-level decomposition described in Section~\ref{ssec:dist_op}. The first level arises from $w$-stacking, where $P$ workers--each corresponding to a $w$-layer--independently perform Fourier transforms. The second level focuses on distributing the Fourier de-gridding/gridding operations enabled by the Fourier-based partitioning of the de-gridding matrices $(\Gb_p)$, 
or the holographic matrices $(\Hb_p)$ when dimensionality reduction is enabled.

When deploying multiple compute nodes using the SLURM scheduler, MATLAB's parallelisation is process-based rather than thread-based, with a one-to-one mapping between workers and CPU cores and no shared memory across workers. Tasks are therefore distributed across all available CPU cores following a horizontal computational hierarchy. Workers are classified in three categories depending on their roles, with some executing multiple tasks sequentially. First, the master worker hosts both the dirty image $\xb_{\textrm{dirty}}$ and the current image estimate ${\xb}^{(k)}$ at a given iteration $k$ of the FB algorithm. Second, FFT workers apply the image-domain $w$-modulation and perform the Fourier transforms (i.e.~the operator $\Fb \Zb_p$ and its adjoint), with the master worker also serving as the FFT worker for the first $w$-layer. Third, the de-gridding/gridding workers host the sparse interpolation matrix blocks $\Gb_{p,c}$ and apply them together with their adjoint (i.e.~complex conjugate transpose), or host and apply the holographic matrix blocks $\Hb_{p,c}$ when dimensionality reduction is enabled. In principle, all workers can serve as de-gridding/gridding workers, as long as sufficient memory buffer is available for their intended task. In particular, FFT workers may handle smaller blocks of the encoded sparse matrix than de-gridding/gridding workers when FFTs are applied to large images. For wideband imaging, a dedicated pool of workers is assigned to each observation frequency following the same allocation of tasks.

The application of $\Phib^\dagger \Phib$ to compute the image $\text{Re}\{\Phib^\dagger \Phib {\xb}^{(k)}\}$ from an image estimate ${\xb}^{(k)}$ is summarised as follows. First, the master worker sends copies of ${\xb}^{(k)}$ to FFT workers. The latter send specific Fourier coefficients to their corresponding de-gridding/gridding workers. These workers compute the gridded model visibilities and send them back to FFT workers. The real parts of the resulting images $\Phib_p^\dagger \Phib_p {\xb}^{(k)}$ are gathered from all FFT workers and summed at the master worker.

{Cross-worker communication, whether involving (i) the transfer of image-sized variables to and from FFT workers or (ii) the transfer of Fourier coefficients to and from de-gridding/gridding workers, is performed using a binary tree communication pattern, in which data are propagated hierarchically to reduce communication overhead and improve scalability. In this structure, workers are organised as nodes of a tree, with each worker communicating only with its parent and up to two child workers. Furthermore, the images $\text{Re}\{{\Phib_{p}^\dagger \Phib_{p}}{\xb}^{(k)}\}$ are progressively summed along the tree, thereby avoiding a summation bottleneck at the master worker which hosts the final image $\text{Re}\{{\Phib^\dagger \Phib}{\xb}^{(k)} \}= \sum_{p=1}^{P} \text{Re}\{{\Phib_p^\dagger \Phib_p}{\xb}^{(k)}\}$.}

{For on-the-fly computation and application of $\Phib^\dagger\Phib$, specifically during the Fourier de-gridding/gridding operations, all workers are deployed in computing and applying the de-gridding matrix blocks through dynamic load balancing. The resulting gridded model visibilities are then gathered at the corresponding FFT workers.}

\subsection{\texorpdfstring{$\Phib^\dagger \Phib$}{phidagphi} computation \& application workflow}

The implementation of $\Phib^\dagger \Phib$ follows a three- to four-step workflow, as illustrated in Figure~\ref{fig:workflow}. First, the compute resource planning stage uses a small subset of the $(u,v)$-coverage to determine the $\Phib^\dagger \Phib$ computation strategy: either precomputed, if the system's memory budget can accommodate its underpinning sparse matrix (potentially requiring data dimensionality reduction), or otherwise on-the-fly. The number of $w$-layers $P$ is also determined accordingly.

When $\Phib^\dagger \Phib$ is precomputed, the second stage consists in memory-controlled Fourier partitioning of the full $(u,v)$-coverage. This uses logical $w$-projection kernels to estimate the memory footprint of the sparse matrix blocks. During this stage, the allocation of CPU cores to each $w$-layer is also determined. The third stage consists in precomputing the sparse (de-gridding or holographic) matrix blocks. The resulting matrix blocks remain accessible on the corresponding workers. The fourth stage corresponds to the repeated application of the precomputed operator within the imaging algorithm.

When on-the-fly computation is considered due to limited compute resources, Fourier partitioning is performed without memory constraints enabling parallel distributed computation of the de-gridding matrix blocks within the imaging algorithm.

%-------------------------------------------------
\section{MATLAB code library}\label{sec:matlab}
In this section, we describe the modular architecture of the MATLAB code library of the RI imaging framework encompassing the proposed measurement operator model. The framework is inherently modular, a direct consequence of the FB iterative structure underpinning the supported imaging algorithms. It comprises a shared data-fidelity module, and an image-regularisation module, which incorporates both monochromatic and wideband denoisers associated with the different imaging algorithms.

\paragraph*{Data fidelity module.}~It is composed of multiple MATLAB classes dedicated to computing and applying the measurement operator and its adjoint. The module supports a CPU-based implementation. In what follows, we describe its main classes. The first class handles the reading of key observational settings including the coordinates $(u,v,w)$, and identifies the $w$-layers ($w^{\textrm{stack}}_p$) via $k$-means clustering and the resulting $w$-offsets. The second class performs parallel memory-controlled Fourier partitioning of the $(u,v)$-coverage both across and within the $w$-layers, and re-orders the associated $(u,v,w)$ coordinates and data vectors accordingly. The third class computes in parallel the sparse matrix blocks of $\Phib^\dagger\Phib$ using an SPMD (single program multiple data) pool, where all available workers are deployed with dynamic load balancing to maximise the use of the available compute resources. The dirty image is computed once and is accessible in the main worker. The operator of interest $\Phib^\dagger \Phib$, and when $\Gb$ is encoded, the operators $\Phib$, and $\Phib^\dagger$ are defined using anonymous functions in MATLAB. 

\paragraph*{Image regularisation module.}~It comprises functions associated with the denoisers, specifically the proximal operators of optimisation algorithms SARA and Hyper-uSARA, and the DNN denoisers of the PnP algorithms AIRI and HyperAIRI. The module supports a CPU-based implementation for SARA-based denoisers, and a GPU-based implementation for AIRI-based denoisers. All denoisers are shipped with spatial faceting functionality for distributed and parallel processing across multiple workers, ensuring scalability to large image sizes. For optimisation algorithms, a faceted implementation of the SARA dictionary is deployed, enabled by its convolutional structure and the compact support of the wavelet kernels \citep{Prusa2012}. For PnP algorithms, the DNN denoiser is applied in parallel on image facets, enabled by the small receptive field of its convolutional layers.

\paragraph*{Main scripts.}~ The library provides two main scripts. The first, \texttt{planner.m}, implements the planning stage and may be executed independently. The second, \texttt{imager.m}, executes the remaining stages of the workflow and performs image formation using the iterative imaging algorithm of choice. First, Fourier partitioning is performed either serially across the $w$-layers using a local \texttt{parcluster} profile or in parallel via MATLAB Batch processing using a SLURM-backed \texttt{parcluster} configuration. Second, the subsequent precomputation of $\Phib^\dagger\Phib$ is carried out within an SPMD pool, which may be launched either locally (when sufficient CPU resources are available) or through a SLURM-backed configuration for on-demand CPU allocation. The SPMD pool remains active throughout the iterative imaging algorithm to support repeated application of the operator. In the on-the-fly mode, $\Phib^\dagger\Phib$ is evaluated at each iteration within an SPMD pool.

\paragraph*{Input \& ouput files.}~Both scripts can take as input the same \texttt{.yml} file containing the complete list of parameters required for the different stages of the workflow. Both also require a \texttt{.mat} observation specification file containing the $(u,v,w)$ coordinates and the details of the observation frequency band. Specific to \texttt{imager.m}, one data file per observation frequency is required, containing the measured visibilities, a noise standard deviation vector, and a binary flag vector indicating measurements to be discarded. In the wideband setting, a separate data file is expected for each observation frequency. Python scripts based on the MeqTrees library \citep{Noordam2010} are provided to convert standard RI measurement sets into MATLAB-compatible files. Output files consist of the estimated model image and the associated residual dirty image, both saved in \texttt{.fits} format.

%-------------------------------------------------
\section{Validation} \label{sec:sim}
We evaluate the distributed resource-adaptive implementation of the operator $\Phib^\dagger \Phib$ in a large-scale widefield observation setting, representative of RI observations with modern telescopes. The analysis does not address the performance of the framework in image formation {in terms of imaging quality}, as thorough validation on monochromatic and wideband real observations with MeerKAT and ASKAP has been conducted in previous works \citep{dabbech22, wilber23a, wilber23b, tang26}, as summarised in Section~\ref{ssec:realdata}.

\begin{figure*}
 \centering
\includegraphics[width=1\linewidth]{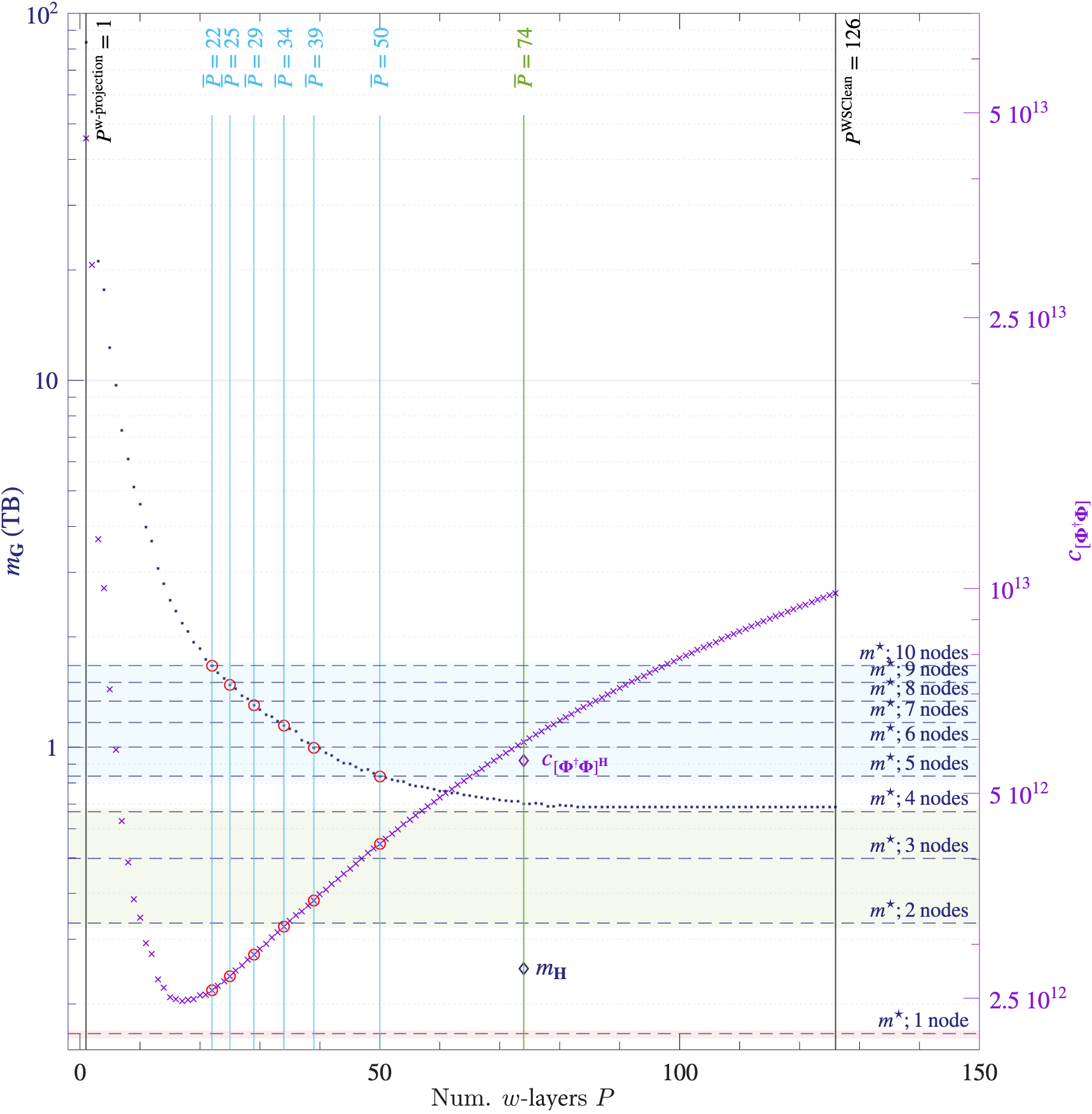}
 \caption{{Compute resource planning: Estimated computational complexity of $\Phib^\dagger \Phib$ (${c}_{[\Phib^\dagger \Phib]}$; right $y$-axis) and the memory required to store the underpinning sparse matrix $\Gb$ ($m_{\Gb}$; left $y$-axis) as a function of the number of $w$-layers ($P$). Dashed horizontal lines indicate the system's memory budget ($m^\star$) for varying number of CPU nodes ranging from 1 to 10. Vertical lines (in blue and green) indicate the corresponding optimal number of $w$-layers $\overline{P}$. For reference, the number of $w$-layers is also shown for \textit{pure} $w$-projection ($P^{w\textrm{-projection}} = 1$) as well as \textit{pure} $w$-stacking using the recommended value in WSClean ($P^{\textrm{WSClean}} = 126$, following \eqref{eq:wplanescondt}). The shaded areas indicate the adopted strategy for computing $\Phib^\dagger \Phib$: (a, blue) precomputing it using $\Gb$, (b, green) precomputing it using $\Hb$ ($\overline{P} = 74$; corresponding memory requirement $m_{\Hb}$ and computational complexity ${c}_{[\Phib^\dagger \Phib]^{\Hb}}$ are also shown), and (c, red) computing it on the fly ($\overline{P} = 74$).}}
 \label{fig:planning}
\end{figure*}

\subsection{Experimental setup}
We simulate MeerKAT data at L-band following the observational setup of real measurements of the Norma cluster \citep{Ramatsoku20} imaged in our previous work using uSARA and AIRI \citep{dabbech22}.
All 64 antennas of MeerKAT are assumed operating over a total duration of 7.8 hours, at the frequency range $960$--$1180$ MHz and with a channel width of 836 kHz. This configuration yields 220 frequency channels, each containing $5.38\times10^6$ visibilities, for a total of $M = 1.18\times10^9$, corresponding to approximately 18 gigabytes (GB) of data in double precision. For the ground-truth intensity image, we consider a simulated continuum image from the SKA Science Data Challenge 1 \citep{sdc1}, which provides a realistic sky model. The original image is of size $32000\times32000$ pixels at 1.4 GHz, generated from a total integration time of 1000 hours. In this work, we take the central $8192\times8192$ pixel region as the ground truth. The pixel size is set to 1.5 arcsec, resulting in a field of view of 3.41 degrees. Finally, the ratio between the spatial Fourier bandwidth of the ground truth and that of the acquisition--determined by the maximum projected baseline at the highest observation frequency--is 2.34. {For reference, {pure} $w$-stacking as implemented in WSClean suggests $P^{\textrm{WSClean}} = 126$. This value can be seen as an upper-bound on the considered number of $w$-layers in our framework. }

Experiments are conducted on the high-performance computing (HPC) facility Cirrus\footnote{\href{http://www.cirrus.ac.uk}{http://www.cirrus.ac.uk}. Cirrus underwent a complete upgrade in December 2025 after the experiments reported in this work were conducted.}. It is an SGI ICE XA system composed of 280 CPU nodes, each with two 2.1 GHz 18-core Intel Xeon E5-2695 (Broadwell) series processors, and with 256 GB of shared memory. The system has a single Infiniband FDR network connecting nodes with a bandwidth of 54.5 GB per second.

\subsection{Compute resource planning} 
We first showcase the proposed resource-adaptive planning strategy under different configurations of the compute resources by varying the number of CPU nodes assumed available from 1 to 10. We recall that each CPU node comprises 36 CPU cores with a total memory of 256 GB, corresponding to approximately 7 GB of memory available per core. The system's memory budget $m^\star$ for hosting the sparse matrix underpinning $\Phib^\dagger \Phib$ is set to about $65\%$ of the memory available per CPU core, reserving a buffer memory for the other operations involved.
To estimate $\Phib^\dagger \Phib$ computational complexity $c_{[\Phib^\dagger\Phib]}$ and the memory footprint of its de-gridding matrix $m_{\Gb}$, we use $1\%$ of the data. When the investigation of dimensionality reduction is triggered for memory reasons, we instead use the full data to obtain a more accurate estimate of the holographic matrix memory footprint $m_{\Hb}$, which may be significantly underestimated when using a small subset of the data.

Figure~\ref{fig:planning} exemplifies different scenarios triggering the three strategies for computing $\Phib^\dagger \Phib$ which are identified during resource planning. Case (a) corresponds to precomputing $\Phib^\dagger \Phib$ by encoding $\Gb$ and is activated when the number of CPU nodes exceeds 4. The selected number of $w$-layers $\overline{P}$ decreases with the increase in memory budget. Case (b) corresponds to precomputing $\Phib^\dagger \Phib$ by enabling dimensionality reduction, where $\Hb$ is encoded at a fixed value $\overline{P}=74$, and is activated when the number of CPU nodes is between 2 and 4. Case (c) arises when only 1 CPU node is available, where the operator is computed and applied on the fly without explicit storing of its sparse matrix with $\overline{P}=74$.

\begin{figure}
 \centering
\includegraphics[width=1\linewidth]{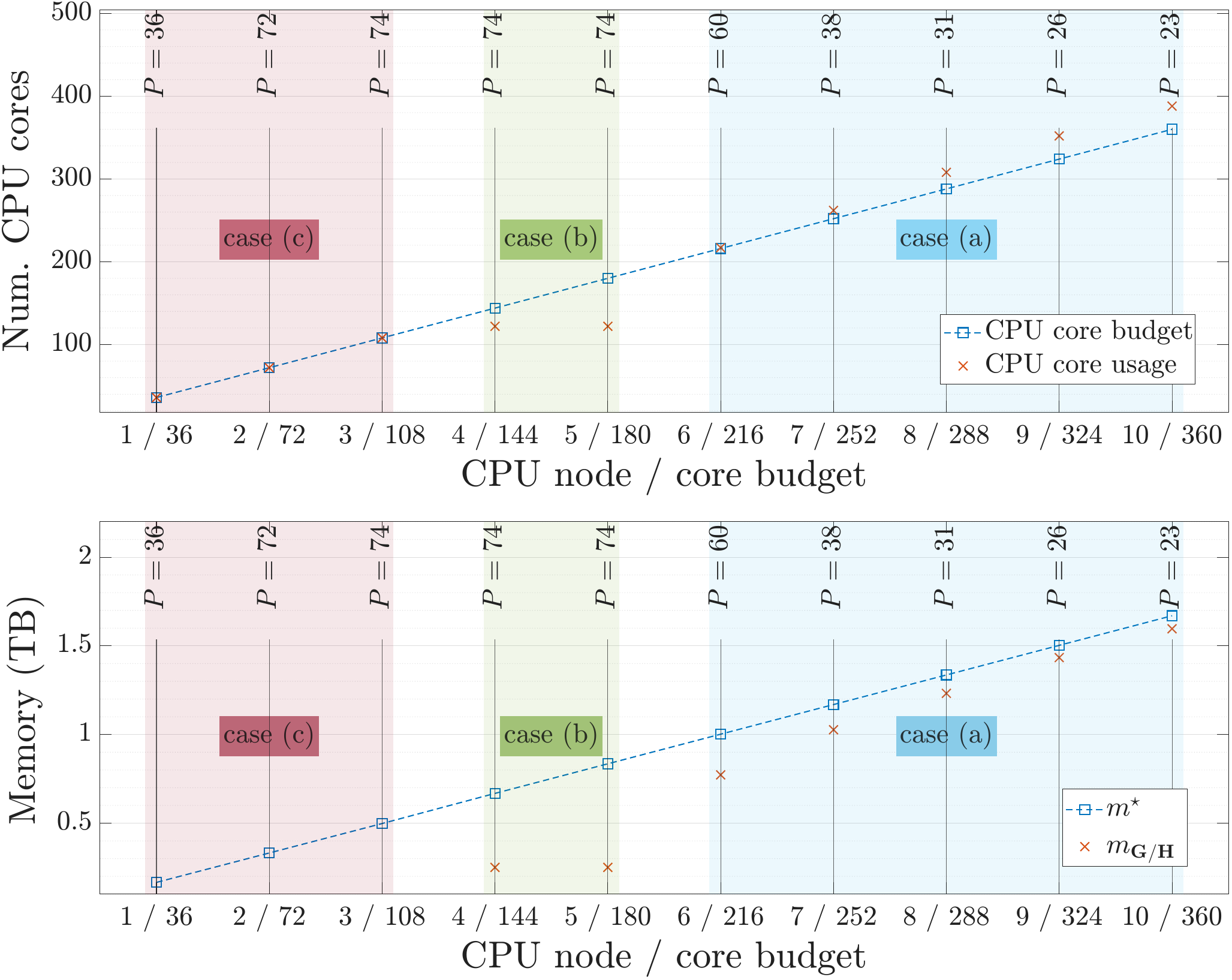}
 \caption{{Compute resource planning as a function of CPU node budget, equivalently expressed in CPU core budget (1 CPU node comprises 36 cores). The selected number of $w$-layers ${P}$ in each setting is shown. The coloured regions indicate the strategy for computing $\Phib^\dagger \Phib$ (see Figure~\ref{fig:decisiontree}): (case (a), blue) precomputing it using $\Gb$, (case (b), green) precomputing it using $\Hb$ (i.e.~data dimensionality enabled), or (case (c), red) computing it on the fly. Top: CPU nodes usage, determined during Fourier partitioning stage, versus CPU node budget, considered at the planning stage. Bottom: the memory footprint of $\Gb$ ($m_{\Gb}$; case (a)) or $\Hb$ ($m_{\Hb}$; case (b)) versus the memory budget $m^\star$. On-the-fly computation does not require storing any of the sparse matrices.}} 
 \label{fig:adj_planning}
\end{figure}

Adjustments to the selected value of $P$ from the optimal value $\overline{P}$, as well as the strategy adopted to compute $\Phib^\dagger \Phib$ may be required to remain within the available CPU-core budget (i.e.~number of workers). Representative examples are shown in the top panel of Figure~\ref{fig:adj_planning}. Such adjustments arise, for instance, when the number of available workers is smaller than the selected number of $w$-layers, as observed for the configuration with 2 CPU nodes, where only 72 CPU cores are available while $\overline{P}=74$. Another example occurs when the number of workers required to accommodate the matrix $\Hb$ exceeds the available resources. This is the case for the configuration with 3 CPU nodes, where 108 CPU cores are available whereas 122 workers are required. In both cases, the decision tree switches to on-the-fly computation of the operator. In the former case, the value of $P$ is reduced to match the number of available CPU cores $(P=72)$. In the latter case, the selected value remains unchanged $(P=\overline{P}=74)$. {In fact, increasing it to match the number of workers would increase the computational complexity of $\Phib^\dagger \Phib$ due to the additional FFTs, whereas the matrix-vector multiplication cost associated with $\Gb$ remains unchanged since its memory footprint is already stable. Nonetheless, on-the-fly computation and application of $\Gb$ leverage all available CPU cores, with workload evenly distributed across workers.}

Additional cases require a slight increase in $P$ to account for a potential extra edge-worker allocation per $w$-layer during the Fourier-partitioning stage, since memory requirements are estimated cumulatively across all $w$-layers. Increasing $P$ reduces the memory footprint of $\Gb$ sufficiently to ensure that enough CPU cores remain available to accommodate the sparse matrix while also reserving one additional worker per $w$-layer. This is applied to configurations with 6 CPU nodes or more. When this is not feasible, dimensionality reduction is enabled, as in the configuration with 5 CPU nodes.

Although these scenarios are identified during the planning stage, additional CPU cores may still be required at the Fourier partitioning stage since planning relies on only a small subset of the data. This behaviour is observed for configurations with 7 or more CPU nodes, where the number of deployed CPU cores, determined by the Fourier partitioning stage, exceeds the planned budget by up to $10\%$. An automated strategy for selecting the size of the data subset used in the compute resource planning stage is therefore needed to mitigate this risk. As for memory, the bottom panel of Figure~\ref{fig:adj_planning} shows that the footprint of the precomputed sparse matrix remains within the memory budget considered during planning for all considered configurations.

\begin{figure}
 \centering
 \includegraphics[width=1\linewidth]{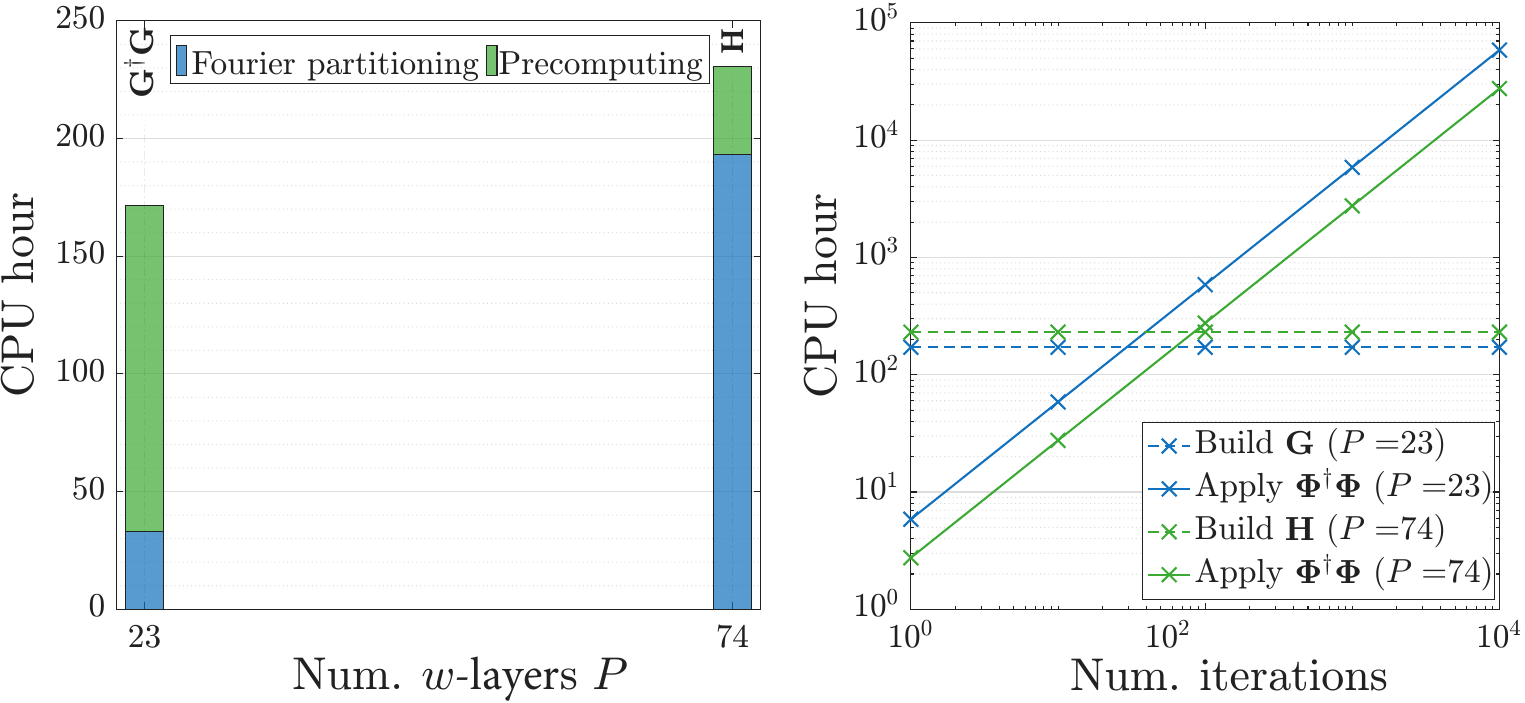}
 \caption{Computational cost of building $\Phib^\dagger \Phib$, specifically its underpinning sparse matrix (including Fourier partitioning and precomputation), and its application within an iterative imaging algorithm, in CPU hour. Results are shown for two parameter choices for the implementation of the precomputed $\Phib^\dagger \Phib$: (i) using $\Gb$ with $P=23$, the closest value to the minimiser of the computational complexity, and (ii) using $\Hb$ with $P=74$, where data dimensionality reduction is enabled. Left: the cost to build the operator as a function of $P$. 
 Right: the cost to build the operator versus the cumulative cost of its repeated application as a function of the number of iterations.} 
 \label{fig:cost}
\end{figure}

\begin{figure}
 \centering
\includegraphics[width=1\linewidth]{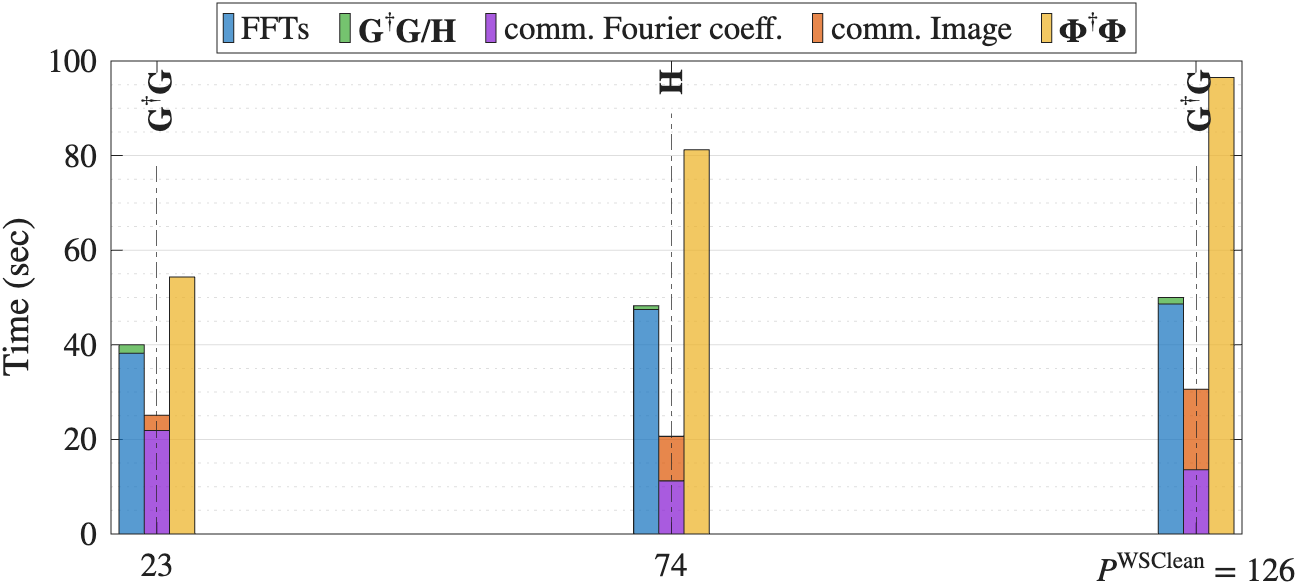}
 \caption{{Breakdown of the computational and communication time associated with the application of $\Phib^\dagger \Phib$ to an image estimate $\xb$ as a function of the number of $w$-layers ${P}$, in seconds (sec). 
 Results are shown for three implementations of the precomputed $\Phib^\dagger \Phib$: (i) using $\Gb$ with $P=23$, the closest value to the minimiser of the computational complexity, (ii) using $\Gb$ with $P=P^{\text{WSClean}}=126$, representing an upper bound on the number of $w$-layers, (iii) using $\Hb$ with $P=74$, where data dimensionality reduction is enabled. The computational time accounts for FFTs and $w$-stacking operations, as well as Fourier de-gridding/gridding operations implemented via either $\Gb^\dagger \Gb$ or $\Hb$. The communication time accounts for cross-worker exchange of the image estimate to and from FFT workers, and transfer of Fourier coefficients to and from de-gridding/gridding workers.
 }}
 \label{fig:time}
\end{figure}

\subsection{\texorpdfstring{$\Phib^\dagger \Phib$}{phidagphi} computation \& application}
We evaluate the computational cost of $\Phib^\dagger \Phib$ from Fourier partitioning and precomputation of the underpinning sparse matrix, to its repeated application within an iterative imaging algorithm, displayed in Figure~\ref{fig:cost}. We consider two parameter choices for the implementation of $\Phib^\dagger \Phib$ identified during the planning stage under the respective compute-resource configurations of 10 and 4 CPU nodes: (i) precomputation using $\Gb$ with $P=23$ corresponding to closest value to the minimiser of computational complexity; and (ii) precomputation using $\Hb$ with $P=74$. {Under these considerations, data dimensionality reduction enabled a 7-fold reduction in memory requirements to store $\Hb$ compared to $\Gb$.

The cost of the Fourier partitioning and the precomputation of the sparse matrix underpinning $\Phib^\dagger \Phib$ is shown in the left panel of the figure. Their cumulative cost under dimensionality reduction via $\Hb$ is approximately $30\%$ higher than that of $\Gb$, and is primarily dominated by the Fourier partitioning stage, which is more complex in this case. By contrast, the cumulative cost using $\Gb$ is dominated by the computation of the sparse matrix due to its larger size.

We examine the interplay between the cost to build the precomputed $\Phib^\dagger \Phib$ (including both the Fourier partitioning and the precomputation stages) and the cost of its repeated application within an iterative algorithm as a function of the number of iterations, displayed in the right panel of the figure. In both settings, the cumulative cost of applying the operator matches the cost to build it within tens of iterations, and subsequently exceeds it by approximately one order of magnitude within a few hundred iterations. This highlights the benefit of precomputing $\Phib^\dagger \Phib$ for highly iterative algorithms such as AIRI and uSARA, typically requiring thousands of iterations to achieve precision imaging. Furthermore, enabling data dimensionality reduction via $\Hb$ leads to an overall computational cost that is nearly three times lower than that of $\Gb$. 

For a more detailed analysis of $\Phib^\dagger \Phib$ application cost, we report in Figure~\ref{fig:time} the breakdown of computational and communication times (in seconds) as a function of the number of $w$-layers $P$. For reference, we also consider the implementation of $\Phib^\dagger \Phib$ using $\Gb$ with $P = P^{\text{WSClean}} = 126$, representing an upper bound on the number of $w$-layers within our hybrid approach. In general, the total time of $\Phib^\dagger \Phib$ application is dominated by the FFTs due to the large image size considered, whereas the Fourier de-gridding/gridding operations combined are the fastest, and almost negligible.

When using $\Gb^\dagger \Gb$, the computational time of Fourier de-gridding/gridding remains nearly constant across both values of $P$. This showcases weak scaling of the second-level decomposition as the Fourier partitioning maintains an approximately constant workload per worker through comparable sparsity of the distributed matrix blocks (i.e.~with close number of non-zero entries). 
By contrast, FFT computations exhibit an increase in runtime with $P$, despite being executed in parallel within the SMPD pool. This may be attributed to implementation overheads in MATLAB. Importantly, inspecting the total application time of $\Phib^\dagger \Phib$ and its communication cost suggests that using $P=23$ $w$-layers in our hybrid $w$-stacking/$w$-projection approach is more efficient than the WSClean-suggested value $P=126$ for pure $w$-stacking.
In fact, increasing the number of $w$-layers yields increased master--FFT worker exchange of large image-sized variables even though simultaneously reducing communication of Fourier coefficients due to the smaller number of non-zero entries in the encoded sparse matrix. When dimensionality reduction via $\Hb$ is enabled, Fourier de-gridding/gridding operations become faster, since both operations are unified into a single, sparser matrix application. In addition, the overall communication cost is more balanced between the two communication types, as the increase in image-sized variables is mitigated by the higher sparsity of $\Hb$. The overall time of $\Phib^\dagger \Phib$ application is, however, larger, likely due to additional MATLAB overheads.

For on-the-fly implementation of $\Phib^\dagger\Phib$, the cost of its application at each iteration of the imaging algorithm is dominated by the on-the-fly computation of the underlying de-gridding matrix blocks within the Fourier de-gridding/gridding operations, which is typically of the same order as its precomputation cost. This showcases the significant computational overhead of on-the-fly computation of $\Phib^\dagger\Phib$ within highly iterative imaging algorithms.
\begin{figure*}
 \centering
 \begin{tabular}{c@{\hspace{0.91\tabcolsep}}c}
 \includegraphics[width=0.38\textwidth]{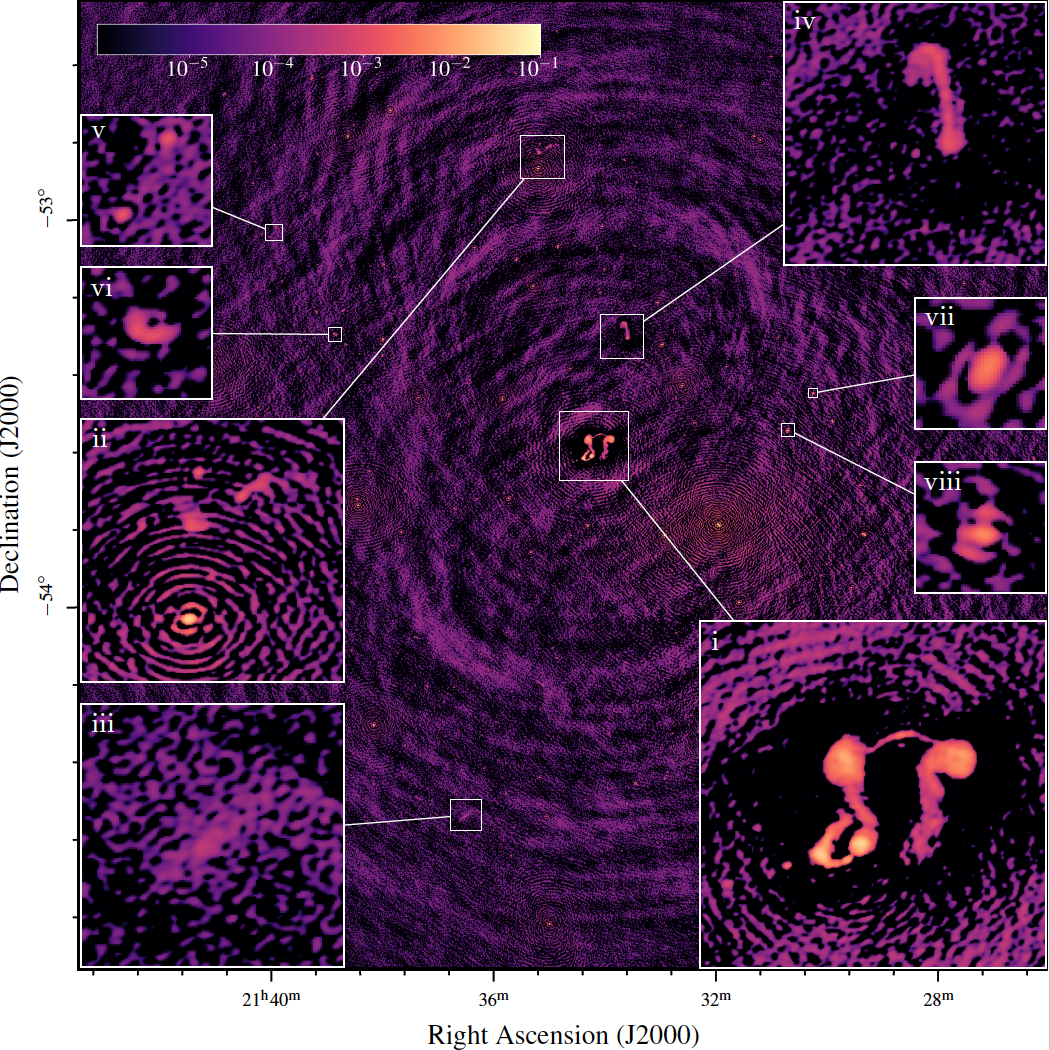}&
 \includegraphics[width=0.38\textwidth]{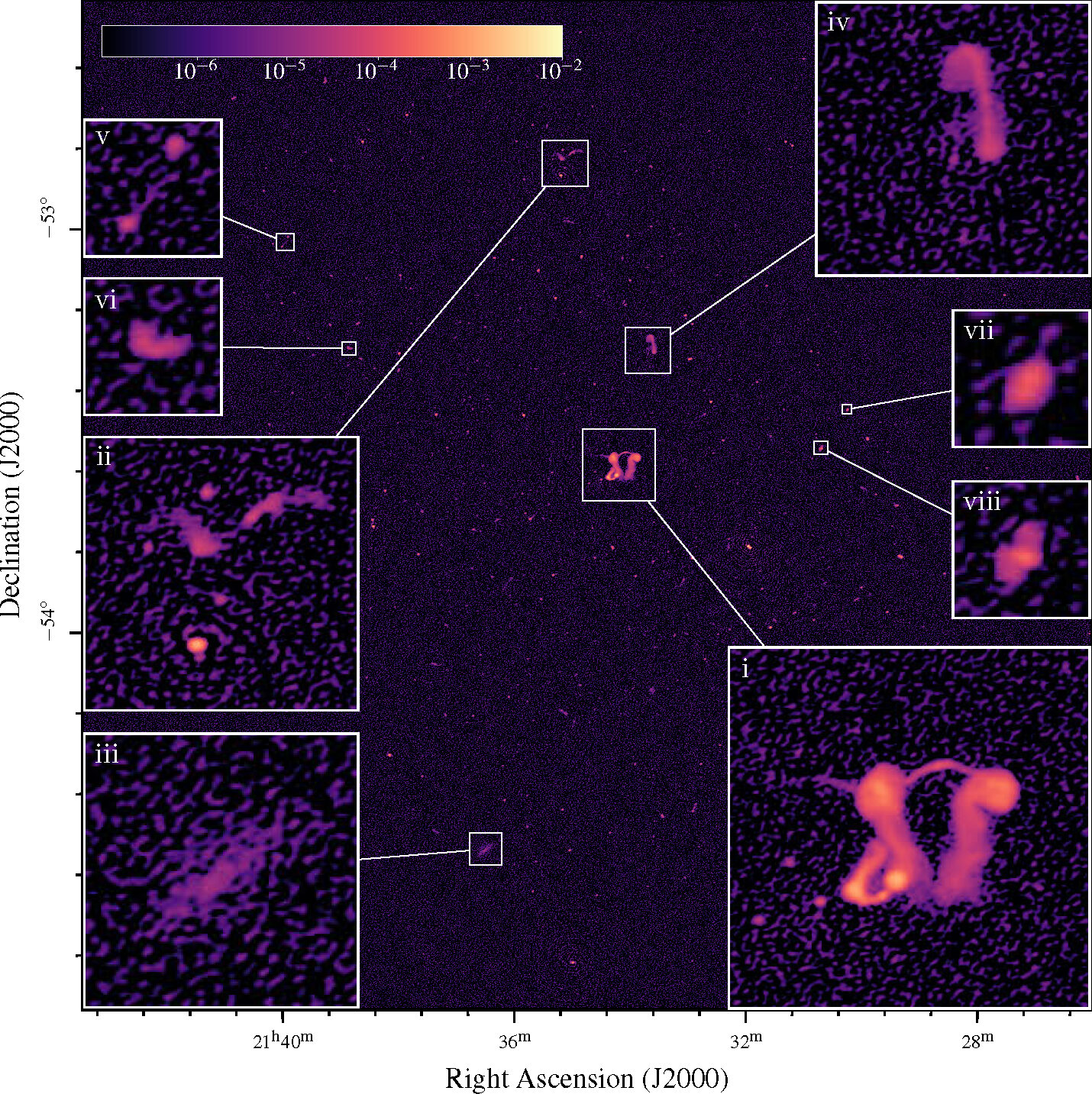}\\
 (a) Dirty image &
 (b) WSClean\\
  \includegraphics[width=0.38\textwidth]{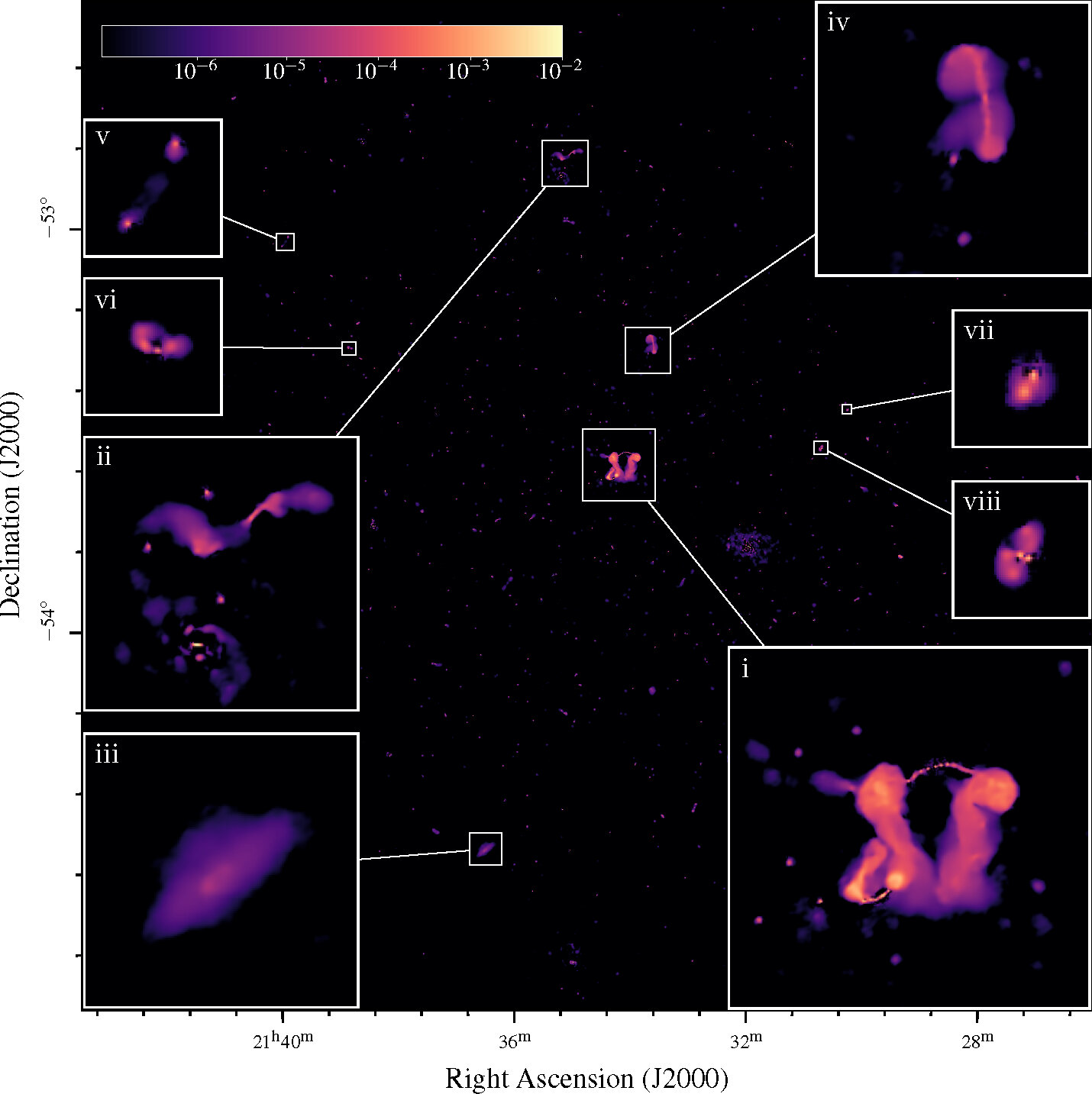}&
 \includegraphics[width=0.38\textwidth]{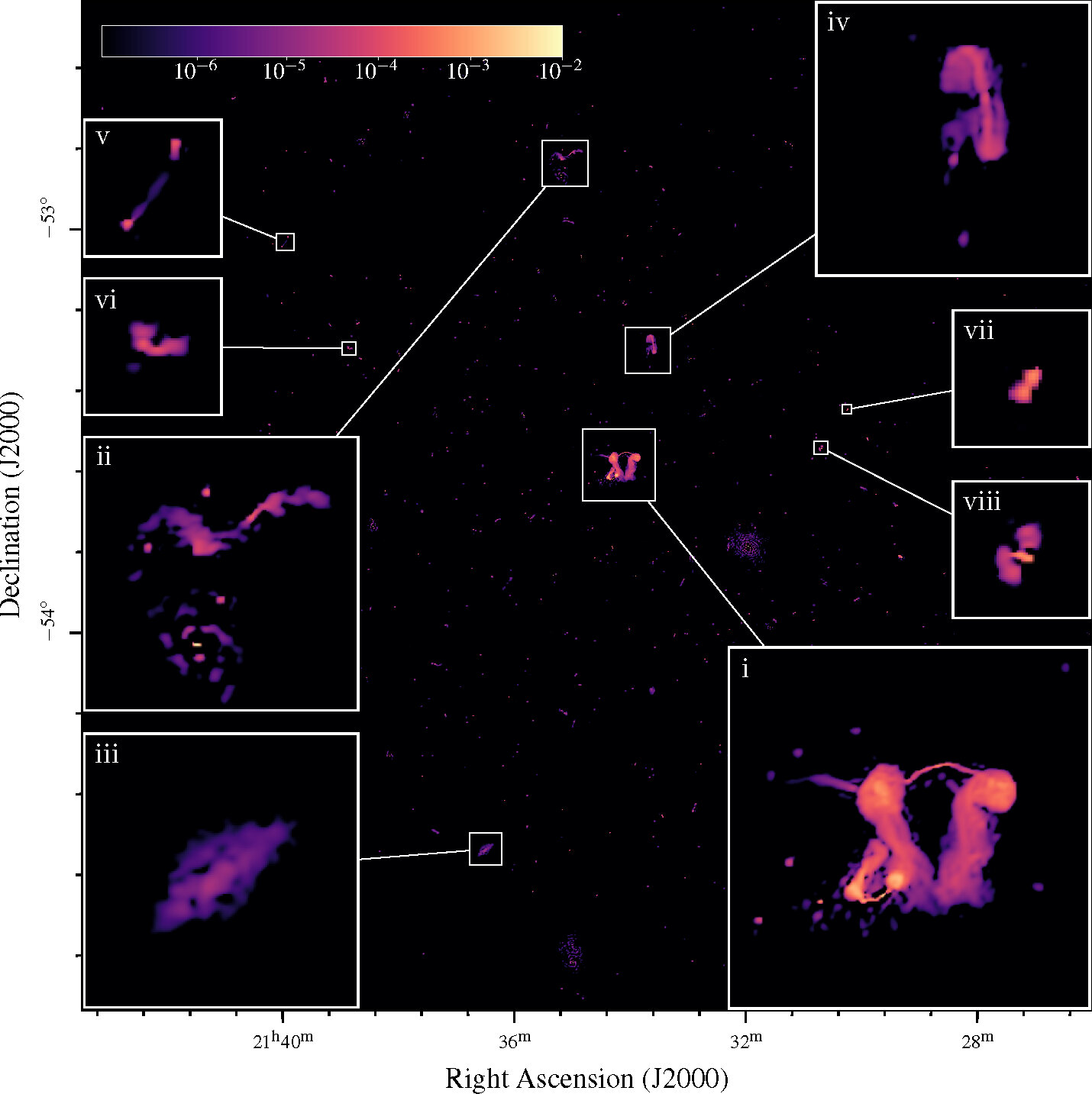}\\
 (c) HyperAIRI &
 (d) Hyper-uSARA \\
 \includegraphics[width=0.38\textwidth]{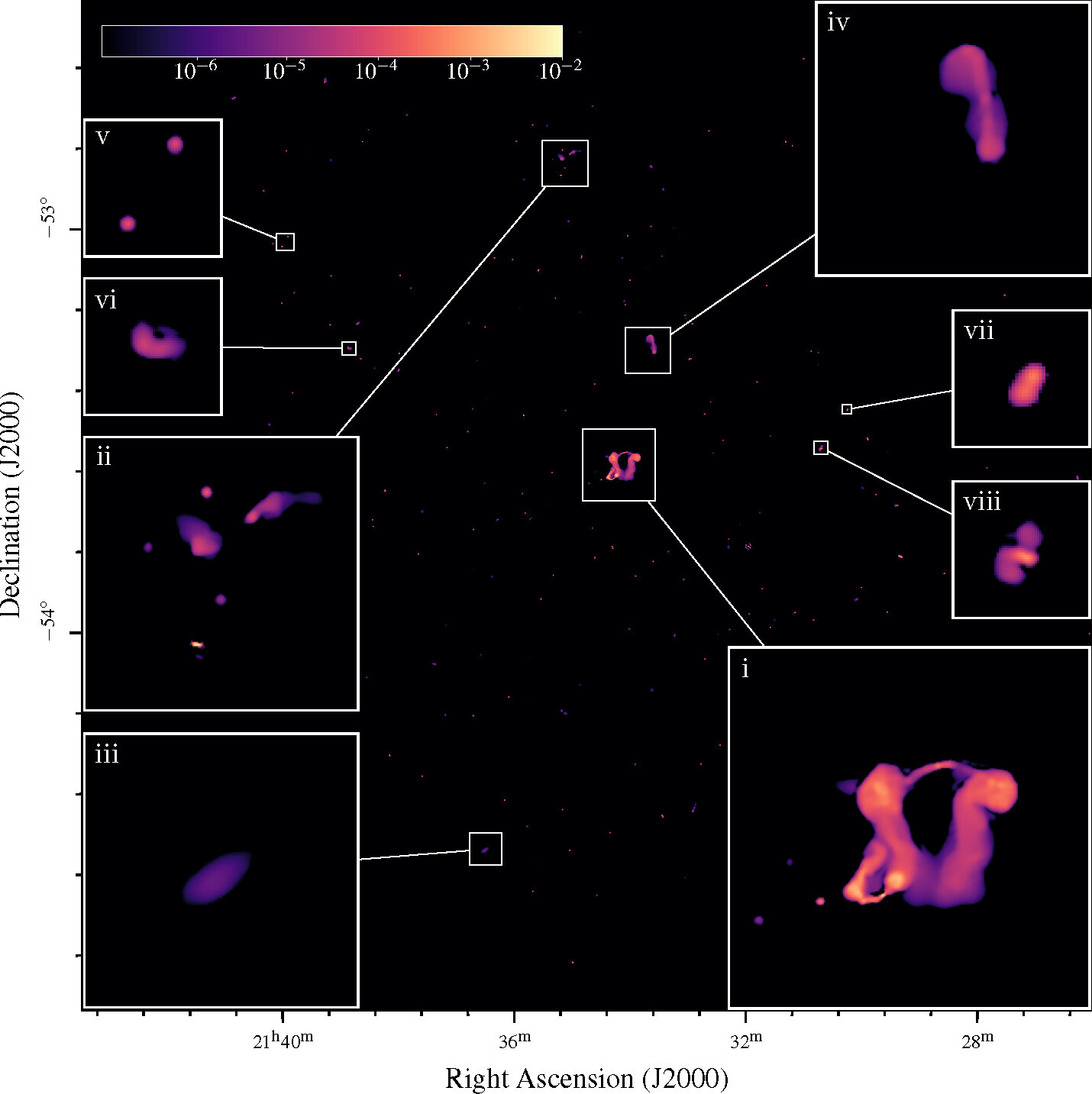} & 
 \includegraphics[width=0.38\textwidth]{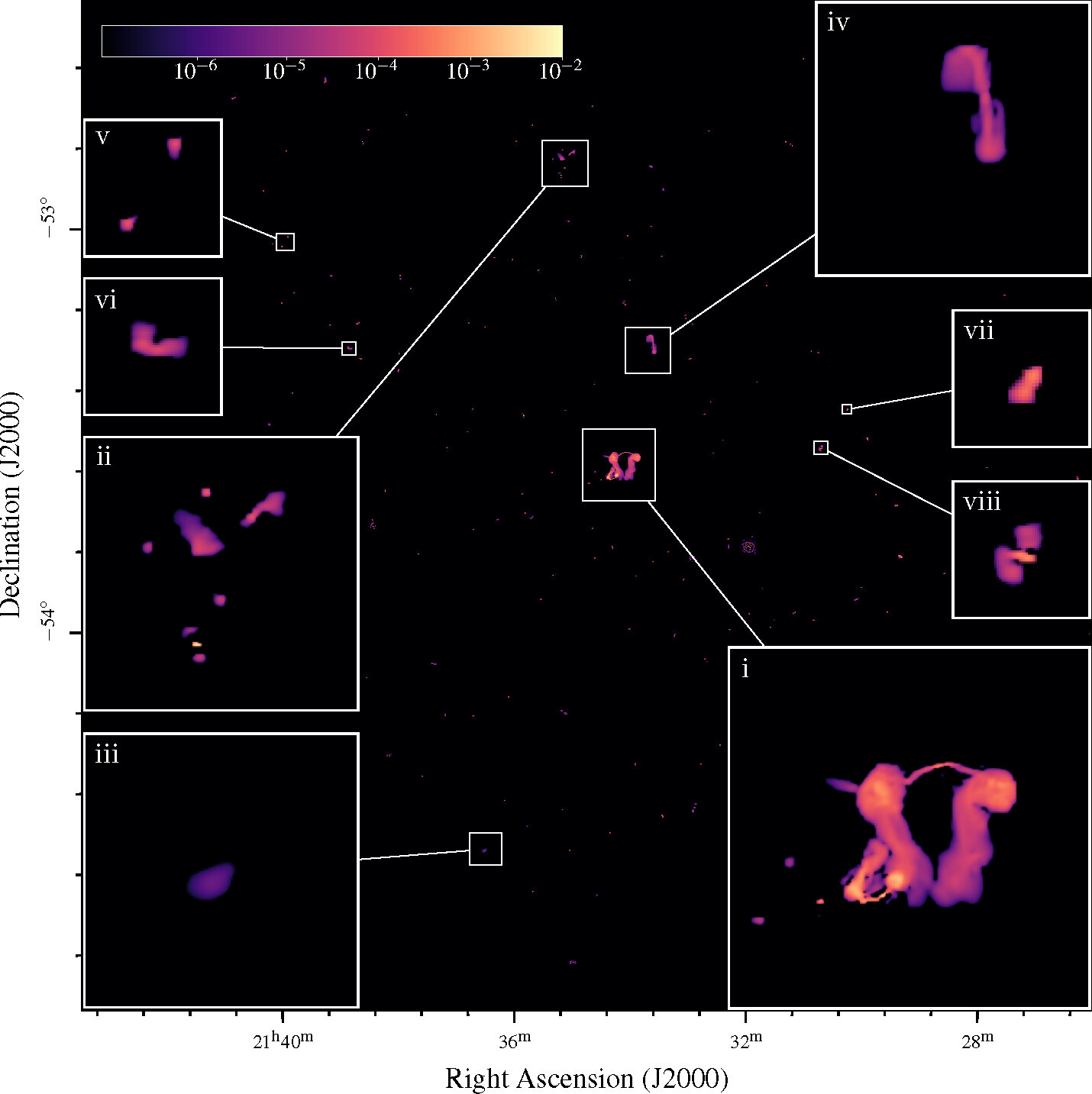}\\
 (e) AIRI &
 (f) uSARA
 \end{tabular}
 \caption{
 {The dirty image (a) of the field SB9442-35 at the frequency 1069 MHz and reconstructed images obtained using WSClean (b) and the four imaging algorithms supported by our proposed framework: HyperAIRI (c), Hyper-uSARA (d), AIRI (e), uSARA (f). Each panel is overlaid with zooms on selected regions: zoom (i) is centred on the ``dancing ghosts''; zooms (ii) and (iv) show regions with both extended and compact radio sources; zoom (iii) contains the star-forming galaxy NGC~7090; zooms (v)-(viii) focus on 4 compact sources. Images taken from previously published results in \citet{tang26}. } } \label{fig:askap}
\end{figure*}

%-------------------------------------------------------------
\section{Real data case studies}\label{sec:realdata}
\label{ssec:realdata}
The proposed framework has been extensively validated on GB-scale real wideband observations for the formation of MB-scale images with both MeerKAT and ASKAP. All experiments were conducted on Cirrus.
Firstly, its data dimensionality feature via the explicit encoding of the holographic matrix has enabled the formation of high-precision 2D radio maps of size $4096 \times 4096$ pixels of the ESO 137-006 radio galaxy in the Norma cluster covering up to 1.91 degrees from about 11 GB of MeerKAT observations, using the monochromatic imaging algorithms uSARA and AIRI \citep{dabbech22}. The number of $w$-layers used is 14, whereas WSClean recommends at least 40 $w$-layers as per the rule \eqref{eq:wplanescondt}. Secondly, its compute resource planning was later thoroughly validated on wideband observations from the ASKAP Early Science and Evolutionary Map of the Universe Pilot surveys, enabling the formation of 2D images of size $5500 \times 5500$ pixels spanning the full FoV of individual ASKAP beams \citep[up to 3.3 degrees each;][]{wilber23a, wilber23b}. Specifically, three fields were imaged, including the merging cluster system Abell 3391-95, the merging cluster SPT-CL 2023-5535, and the ``dancing ghosts''. Most recently, the framework has enabled the formation of a 3D image cube of size $4096\times 4096\times 8$ pixels from ASKAP observations of the ``dancing ghosts'' using wideband algorithms Hyper-uSARA and HyperAIRI \citep{tang26}. The number of selected $w$-layers was fixed to 12 per frequency, whereas the number recommended by WSClean ranges from 48 at the lowest frequency to 73 at the highest. Figure~\ref{fig:askap} presents a preview of the wideband imaging results obtained with the different algorithms supported by our MATLAB-prototyped framework.

When compared to WSClean, the computational cost in CPU hour of the considered imaging algorithms remained within one order of magnitude of that of the CLEAN algorithm. This performance was enabled by the hybrid $w$-stacking/$w$-projection approach and the resource-adaptive implementation of the underlying measurement operator, including the optimised selection of the number of $w$-layers, the use of data dimensionality reduction when required by memory constraints, and the fully distributed implementation of the operator for efficient use of the available compute resources.

%-------------------------------------------------------------
\section{Discussion \& conclusions}\label{sec:conclusions}
{We have proposed a distributed resource-adaptive measurement model leveraging the hybrid $w$-stacking/$w$-projection approach for scalable widefield RI image formation. The model is implemented within an imaging framework, supporting state-of-the-art imaging algorithms ranging from the optimisation-based algorithm SARA to the PnP algorithm AIRI and their wideband extensions. These algorithms iteratively apply the measurement operator and its adjoint through the composite mapping  $\Phib^\dagger \Phib$ which encodes a highly position-dependent PSF.  The resulting framework is prototyped in MATLAB with a CPU-based implementation of its underpinning measurement model. While previous works have already validated the framework on real observations from modern telescopes, we have focused herein on the key scalability features underlying $\Phib^\dagger \Phib$, describing them in more detail and analysing their performance at larger scales in both data and image dimensions.}
 
Firstly, we have shown that a multi-level decomposition of $\Phib^\dagger \Phib$, underpinned by memory-controlled Fourier partitioning, enables efficient distribution and parallelisation of the underlying Fourier de-gridding/gridding operations. Secondly, we provide an optional data dimensionality reduction functionality by embedding the Fourier de-gridding/gridding operations in a holographic matrix, fully blind to the data size. Thirdly, we have devised a planning strategy to automate the selection of the adopted strategy for $\Phib^\dagger \Phib$ computation (including whether to enable dimensionality reduction), and the choice of the corresponding number of $w$-layers, thus adapting $\Phib^\dagger \Phib$ implementation to the system's memory and CPU-core budget while minimising its computational complexity. We have demonstrated the planning strategy under different compute-resource configurations, including practical adjustments required for edge cases.

We recognise that modern CLEAN implementations such as WSClean provide highly efficient on-the-fly implementations of the measurement operator. In addition, the more recent $w$-gridder \citep{Arras21} in WSClean adopts a 3D convolutional de-gridding/gridding with controlled accuracy via a user-defined tolerance parameter, and can operate with fewer $w$-layers than the conservative rule of \citet{Offringa2014}. These developments are effective within the major cycle/minor cycle structure of CLEAN, which requires only a limited number of passes through the data. However, they are not designed for more advanced imaging algorithms at the interface of optimisation theory and deep learning, which are typically highly iterative. So far, these algorithms have seen limited adoption in practice, despite their significantly improved imaging precision over CLEAN.

These limitations motivate the proposed framework, which targets a resource-adaptive implementation of the measurement operator through precomputation of $\Phib^\dagger \Phib$ achieved via precomputation of the underpinning sparse matrices associated with the Fourier de-gridding/gridding operations, thus enabling efficient repeated application during image formation. Although the precomputation cost may exceed that of modern on-the-fly implementations in a single evaluation, our results show that within iterative algorithms this cost is amortised after only a few tens of iterations and becomes substantially lower after a few hundred iterations. Validation on real data further confirms the relevance of $\Phib^\dagger \Phib$ precomputation for highly iterative algorithms such as AIRI and uSARA, whose computational costs remain within one order of magnitude compared to WSClean.

Finally, although the proposed scalability features are currently implemented under MATLAB constraints (e.g.~process-based parallelisation and specific memory limits per worker), the main concepts, specifically, the general resource-adaptive strategy, remain relevant for future Python and GPU-based implementations. That said, GPU architectures introduce a different balance between computation, memory usage, and data communication. In particular, the memory constraints and inter-device communication on GPUs differ substantially from the CPU-based architectures. In this context, the detailed implementation choices of the framework should be revisited, including the relative importance of precomputation versus on-the-fly application of the measurement operator.

\section*{Data Availability}
The MATLAB prototype code of the proposed imaging framework is available in the BASPLib code library on GitHub. BASPLib is developed and maintained by the Biomedical and Astronomical Signal Processing Laboratory (\href{https://basp.site.hw.ac.uk/}{BASP}).

%-------------------------------------------------------------
\begin{acknowledgments}
The authors warmly acknowledge Adrian Jackson for valuable discussions on HPC-related implementation questions. This research was supported by UK Research and Innovation through the EPSRC grant EP/T028270/1, and the STFC grants ST/W000970/1 and UKRI1186. The research used Cirrus, a UK National Tier-2 HPC Service at EPCC funded by the University of Edinburgh and EPSRC (EP/P020267/1). 
\end{acknowledgments}
\software{
  NUFFT \citep{Fessler2003},
  MeqTrees \citep{Noordam2010},
}

%%%%%%%%%%%%%%%%%%%%%%%%%%%%%%%%%%%%%%%%%%%%%%%%%%

\bibliographystyle{aasjournalv7}
\bibliography{eira}

\end{document}